\documentclass[prb,twocolumn,showpacs,superscriptaddress,floatfix]{revtex4}
\usepackage{mathptmx}
\usepackage{amssymb,multirow}
\usepackage{amsmath,amssymb,amsfonts,bm}
\usepackage{graphicx,epsfig,color}
\usepackage{latexsym}
\usepackage{dsfont}
\usepackage{color}
\usepackage{wasysym}
\usepackage{soul}

\begin{document}

\title{Ion-beam nanopatterning of silicon surfaces under
co-deposition of non-silicide-forming impurities}

%
%

\author{B. Moon}
\affiliation{Department of Physics, Sook-Myung Women's
University, Seoul 140-742, Korea}
\affiliation{Department of Physics, Korea University, Seoul
136-701, Korea}
\author{S. Yoo}
\affiliation{Department of Physics, Sook-Myung Women's
University, Seoul 140-742, Korea}
\author{J.-S. Kim}
\email[]{jskim@sm.ac.kr}
\affiliation{Department of Physics, Sook-Myung Women's
University, Seoul 140-742, Korea}
\author{S. J. Kang}
\affiliation{Institute for Basic Science: Center for Functional
Interfaces of Correlated Electron Systems, Seoul National
University, Seoul 19-419, Korea}
\author{J. Mu\~{n}oz-Garc\'{i}a}
\affiliation{Departamento de Matem\'aticas and Grupo
Interdisciplinar de Sistemas Complejos (GISC), Universidad
Carlos III de Madrid, 28911 Legan\'es, Spain}
\author{R. Cuerno}
\email[]{cuerno@math.uc3m.es}
\affiliation{Departamento de Matem\'aticas and Grupo
Interdisciplinar de Sistemas Complejos (GISC), Universidad
Carlos III de Madrid, 28911 Legan\'es, Spain}

\date{Received \today }

\begin{abstract}
We report experiments on surface nanopatterning of Si targets
which are irradiated with 2 keV Ar$^+$ ions impinging at
near-glancing incidence, under concurrent co-deposition of Au
impurities simultaneously extracted from a gold target by the
same ion beam. Previous recent experiments by a number of groups
suggest that silicide formation is a prerequisite for pattern
formation in the presence of metallic impurities. In spite of the fact that Au is known not to form stable
compounds with the Si atoms, ripples nonetheless emerge in our
experiments with nanometric wavelengths and small amplitudes,
and with an orientation that changes with distance to the Au
source. We provide results of sample analysis through Auger
electron and energy-dispersive X-ray spectroscopies for their
space-resolved chemical composition, and through atomic force,
scanning transmission electron, and high-resolution transmission
microscopies for their morphological properties. We discuss
these findings in the light of current continuum models for this
class of systems. The composition of and the dynamics within the near-surface amorphized layer that ensues is expected to play a
relevant role to account for the unexpected formation of these
surface structures.
\end{abstract}

\pacs{
79.20.Rf, 
81.16.Rf, 
68.35.Ct, 
05.45.-a  
}

\maketitle

\section{Introduction}
\label{sec:intro}

Ion beam irradiation employing low energy noble gas ions is a functional tool to obtain nanoscale surface patterns in a diversity of materials. \cite{Munoz-Garcia2009book} Under proper conditions these patterns can be ordered over large lateral distances in comparison to their heights.\cite{Facsko1999} In particular, and due to their technological relevance, much work has been focused on the formation of such surface nanostructures on silicon.\cite{Munoz-Garcia2014} The first experiment reporting the formation of silicon nanopatterns by ion irradiation was published in Ref.\ \onlinecite{Gago2001}, in which 1.2 keV Ar$^+$ ions were employed to form ordered nanodots on a Si substrate. Despite many experiments having shown the formation of a large variety of nanostructures on silicon surfaces, it turned out that most of these results were due to the effect of impurities incorporation.\cite{Munoz-Garcia2014} This is, at least, the case for small angles of incidence as first noted by Ozaydin \textit{et al.}\cite{Ozaydin2005} In this work, the authors brought to light the importance of a continuous metal supply to enhance the pattern formation. Thus, only silicon substrates with metal impurity concentrations above a certain threshold may develop nanostructures at normal incidence.\cite{Zhou2011} This point was confirmed by Madi \textit{et al.}, who showed the absence of surface nanopatterns for Ar$^+$ ions at different energies and clean conditions, for incidence angles smaller than 48$^\circ$.\cite{Madi2011} Above this angle, and up to 80$^\circ$, the formation of perpendicular ripples was reported, with wavevector parallel to the projection of the ion beam.

Recent efforts have explored the morphologies formed under systematic impurity incorporation in order to better understand this process and the intervening physical mechanisms. In this line, in 2008 Hofs\"ass and Zhang proposed a setup combining ion irradiation and atom deposition to study ripple formation on silicon surfaces at 70$^\circ$ and 80$^\circ$ with 5 keV Xe$^+$ ions.\cite{Hofsass2008} This technique, known as surfactant sputtering, consists on locating an adjacent plate with a certain tilt next to the substrate. This experimental geometry allows to modify substrate composition and irradiation conditions, since the plate is co-sputtered during ion bombardment and some material is deposited on the substrate. In this experiment, ripples properties were altered depending on the plate material and the distance to it. For instance, the substrate surface roughness was found to increase with the distance to a gold plate. In contrast to the previous experiment in which ripples are formed for large angles of incidence in the absence of contamination, in Refs.\ \onlinecite{Macko2010} and \onlinecite{Zhang2011} similar setups were employed with smaller angles of incidence and an iron plate perpendicular to the substrate, and allowing for different angular locations, respectively. In these experiments, due to redeposition, metal content is maximum next to the plate location and decreases with the distance to it. The sputtered metallic material (and perhaps reflected ions) influences pattern appearance, which would not be formed without this surfactant sputtering. Using 2 keV Kr$^+$ ions at 30$^\circ$, in Ref.\ \onlinecite{Macko2010} different pattern types were observed depending on the distance to the plate. Similarly, parallel mode ripples and dots far from the plate were observed in Ref.\ \onlinecite{Zhang2011} using 5 keV Xe$^+$ ions at normal incidence. Later works have confirmed the dependence of the topography on the plate distance at low temperatures, $140-440\, \mbox{K}$,\cite{Macko2011} confirming metal silicide formation and its segregation as key mechanisms to induce pattern formation. Silicide-induced patterning was also  recently studied by surfactant sputtering employing different co-deposited materials in Refs.\ \onlinecite{Hofsass2013} and \onlinecite{Engler2014}, in which the requisite of silicide formation to observe silicon nanopatterning was confirmed for 5 keV Xe$^+$ ions at 60$^\circ$ and 2 keV Kr$^+$ ions at 30$^\circ$, respectively. Interestingly, in the former work, no ripple formation was observed for the case of copper and gold contamination, suggesting that only materials leading to silicide phase separation may form patterns on silicon substrates.


Although the mechanisms inducing pattern formation under surfactant sputtering are not yet completely understood, from the previous experimental results it seems convenient to distinguish between two different cases: when selective formation of silicide occurs, and when silicon and impurities do not react chemically. The former case has been addressed theoretically in Ref~\onlinecite{Bradley2013a}, where a mathematical model is formulated to describe the height and composition evolutions for normal incidence bombardment and oblique deposition. It is found that silicide formation and the interaction between the surface composition and morphology are crucial to destabilize the surface. For the case in which silicon and impurities do not react chemically, some theoretical works predict pattern formation when impurity deposition is simultaneous to ion bombardment.\cite{Zhou2010,Bradley2011b,Bradley2012b,Bradley2012a} These models predict, for instance, a pattern instability that can emerge purely as the result of the difference in the sputtering yields for both species\cite{Bradley2012a} when a minimal impurity concentration value is reached.\cite{Bradley2011b} A more extended revision of these models will be provided in Section \ref{sec:models}.

In order to check the possibility of nanopattern production for the case of non-silicide formation, in this paper we have conducted experiments in which Au atoms were co-deposited over a silicon substrate using the surfactant sputtering technique. It is important to stress that gold does not form stable silicide, but a metastable alloy with Si within a wide range of compositions below its eutectic temperature.\cite{Hofsass2013} In the experiment described below, well-defined nanopatterns with different features were found on the Si substrate, indicating that silicide formation is not a necessary condition for pattern formation. Based on our experimental findings, we will try to rationalize the role of Au impurities and how the formation of Si nanopatterns is catalyzed by them.

The paper is organized as follows. Section \ref{sec:exp} is devoted to describe our experimental setup, while results thus obtained are reported in detail in Sec.\ \ref{sec:results}. A rationalization of our observations in the light of current continuum descriptions of the process is contained in Sec.\ \ref{sec:models}. This is followed by a discussion in Sec.\ \ref{sec:disc}. The paper closes by summarizing our main results and conclusions in Sec.\ \ref{sec:concl}.

\section{Experiments}
\label{sec:exp}

\begin{figure}[!t]
\centering
\includegraphics[width=0.79\linewidth]{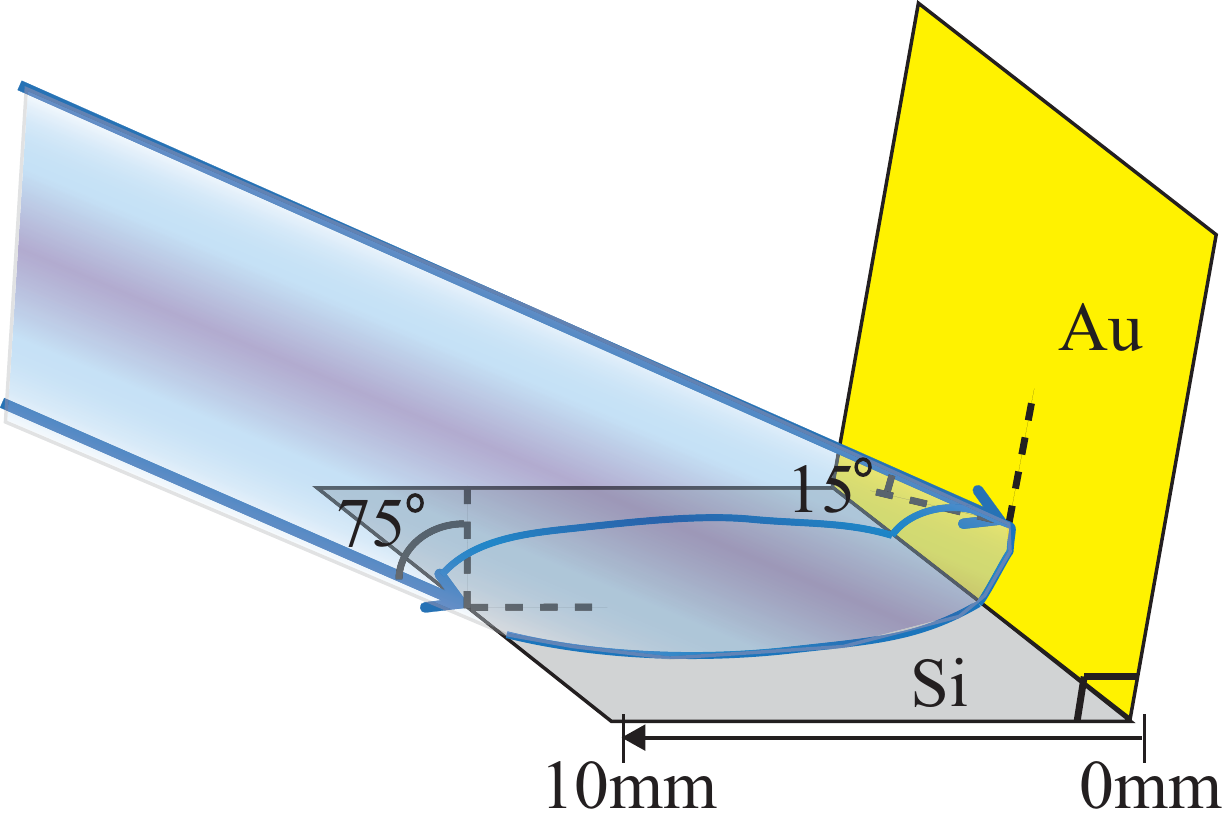}
\caption{Schematic of the geometry employed in this work
for the ion-irradiation of Si(100) substrate with simultaneous
co-sputtering of Au target.
The arrows indicate the Ar$^+$ ion beam.}
\label{Sputter_Geometry}
\end{figure}

Ion-beam sputtering (IBS) has been performed in a custom-built
ultrahigh vacuum chamber with a base pressure of mid-10$^{-10}$
Torr. We have used an (99.999 \%) Ar gas as the ion source. The
ion energy $\varepsilon$ was 2 keV, and the ion flux $f$ was
1.59 ions nm$^{-2}$s$^{-1}$, which is estimated from the target
current. Since secondary electrons are not taken into account,
this is only a nominal value for $f$ which sets an upper limit
for the actual ion flux.
All the images are taken for fixed fluence of 8,586 ions nm$^{-2}$.
We have used a Kauffman-type ion gun
(IQE-11, SPECS), the beam diameter being $ < 10$ mm.

Before loading the sample into the chamber for IBS, 10 mm
$\times$ 10 mm Si(100) chips were immersed into a HF solution
(99\% H$_2$O + 1\% HF) for 5 seconds, in order to remove the
natural oxide from the surface, and then rinsed by de-ionized
water. For the co-deposition of Au during IBS, we mounted the Si
sample at the edge of a single crystalline (99.999\%, Mateck)
Au(001) target at a right angle, using a silicon glue. Thus, the
ion-beam simultaneously irradiated both the Si(100) at a
near-glancing angle of 75$^{\circ}$ and the Au sample at
15$^{\circ}$ from the respective surface normals, see Fig.\
\ref{Sputter_Geometry}.

In order to align the ion beam, we placed two 100 nm-thick Au
films deposited on Si(100) in the same experimental geometry as
shown in Fig.\ \ref{Sputter_Geometry}, and irradiated the films
with the Ar$^+$ ion beam as specified above. The location and
the profile of the beam on the target surfaces is, then,
identified by the erosion profile of the Au films.
That information is used to adjust the beam to be well inside the sample, precluding impurity
deposition from the sample holder, as also confirmed by Auger
electron spectroscopy (AES).

After sputtering the sample, its surface topography was
investigated {\it ex situ} by both an atomic force microscope
(AFM, XE-100, Park Systems) in the tapping mode and a scanning
electron microscope (SEM, JSM-7600F, JEOL). The cross-sectional specimen for the transmission electron microscope (TEM) analysis was prepared using a conventional ion-mill procedure after mechanically grinding the specimen. TEM (JEM 2100F, JEOL) was operated at an acceleration voltage of 200 kV for both the high-resolution (HR-TEM) and scanning transmission electron microscopy (STEM) modes. An annular dark field (ADF) detector ranging from 100 to 250 mrad was used for high angle annular dark field (HAADF) imaging. Energy dispersive X-ray spectroscopy (EDX) analysis was carried out simultaneously with the HAADF-STEM imaging.
Depth profiling was also performed by taking Auger electron
spectra (PHI Nanoprobe 700), while raster-sputtering a 3
mm$\times$3 mm area with the Ar$^{+}$ beam.

\section{Results}
\label{sec:results}

\begin{figure}[!t]
\centering
\includegraphics[width=0.95\linewidth]{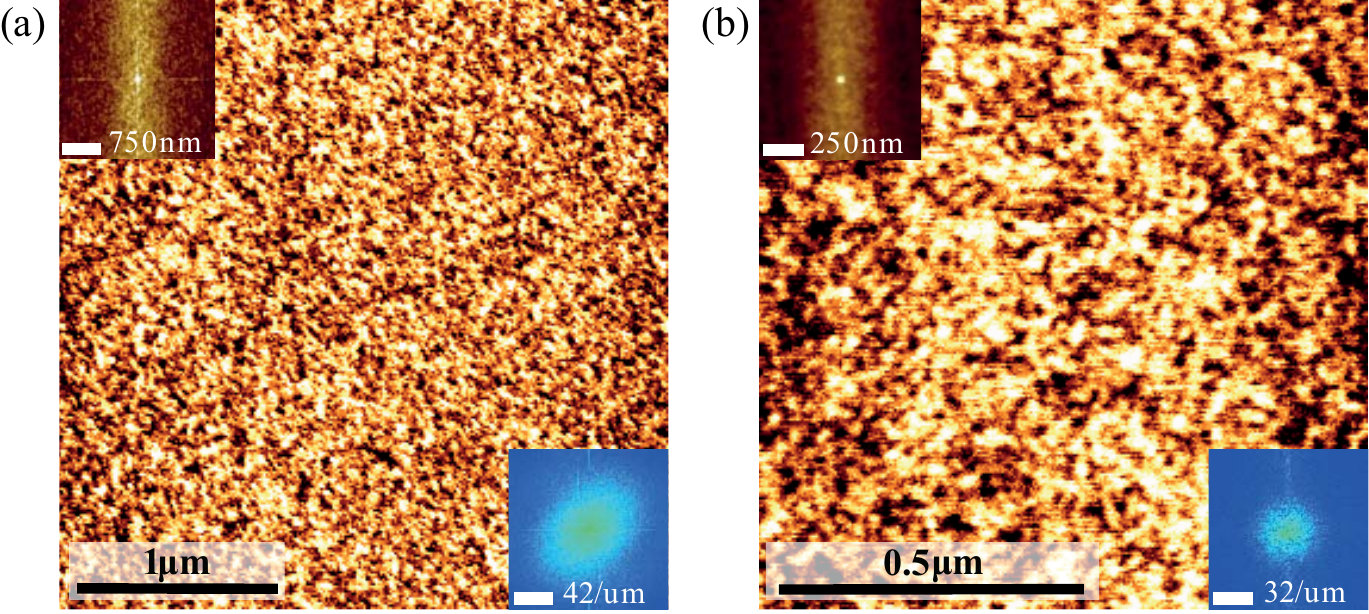}
\caption{Topographies of Si(100) after IBS {\it without} metal
co-deposition at two different lateral scales, 3 $\mu$m (a) and
1 $\mu$m (b). Top-left and bottom-right insets show the corresponding 2D
 auto-correlation and Fourier transform, respectively.}
\label{Null_Pattern}
\end{figure}

Figure\ \ref{Null_Pattern} shows typical (a) 3 $\times 3$
$\mu$m$^2$ and (b) 1 $\times 1$ $\mu$m$^2$ Si surfaces after IBS
under the stated sputtering condition, in the absence of gold codeposition.
No surface pattern is detectable, as suggested by the auto-correlation function and
 2D Fourier transform (FT) provided in the top left and bottom
right insets of each image, respectively. For a height
profile $h(x,y)$, the auto-correlation function is defined as
$\langle h(\mathbf{r_0}) h(\mathbf{r_0}+\mathbf{r})\rangle$,
where $\mathbf{r}=(x,y)$ denotes a point on the target plane and
the brackets denote average with respect to the position of the
reference point $\mathbf{r}_0$. For the sputtered surfaces in Fig.\ \ref{Null_Pattern}, the
values of the surface roughness or width, $W$, are (a) 0.2 nm and (b) 0.1 nm, respectively.
These values are similar to those of the Si substrate prior to irradiation. Here, $W(t)$ is
defined by $W(t)\equiv\sqrt{\langle[h(\textbf{r},t)-\overline{h}(t)]^{2}\rangle}$,
where $\overline{h}(t)$ is the mean height at time $t$. The
bracket denotes the average over the imaged space.

\begin{figure}[!t]
\centering
\includegraphics[width=0.95\linewidth]{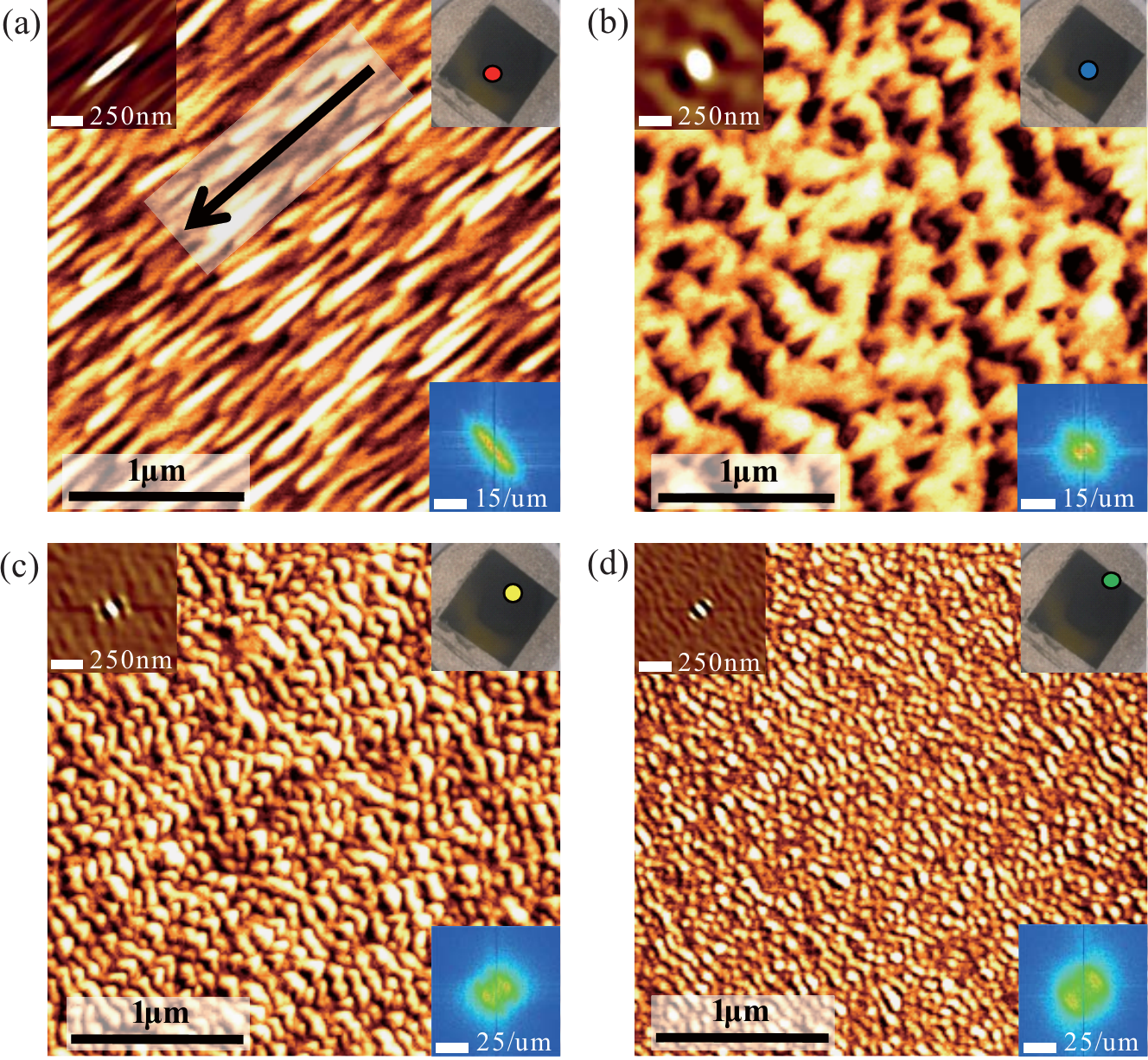}
\caption{The patterns on Si substrate formed by IBS with Au
co-sputtering. From (a) to (d), the distance from the Au source
to the respective imaging site increases as indicated in the top-right inset. (a) Site 1 is in the
{\it relatively} Au-rich region; (b) Site 2 is at a transition
region in which the ripple reorientation is under way; (c) Site
3 is located in a Si-rich region; (d) Site 4 is at the most
Si-rich region near the edge of the Si substrate. The arrow in
(a) indicates the projection of the ion beam direction,
which is common to the four images. Top-left and bottom-right insets show the corresponding 2D
auto-correlation and Fourier transform, respectively. }
\label{Patterns}
\end{figure}

Figure\ \ref{Patterns} presents the surface topographies
obtained at four different sites on the Si(100) target after IBS
with concurrent Au co-deposition. In sharp contrast to the case
of clean Si(100), now well-defined nanopatterns form.
Note that the pattern changes from (a) to (d) in Fig.\
\ref{Patterns} as the imaging site is further away from the Au
source and the flux of co-deposited Au impurities becomes smaller.

Figure \ref{Patterns}(a) shows a ripple pattern near the Au
source (site 1), as indicated in the upper-right inset. As seen
in the real-space image and consistent with the 2D
autocorrelation and FT provided in the insets, the ridges of
the ripples run along the projection of the ion beam direction,
their wavelength being close to 120 nm.
This is the same ripple orientation as obtained on pure Au(001)
targets under a similar sputtering condition.\cite{Kim2009} The
mean uninterrupted ridge length of the ripples ($> 1 \mu$m), or
coherence length, is, nevertheless, much shorter than that for
Au(001). Ripple growth in this {\it relatively} Au-rich region
is probably interrupted by chemical and structural defects,
such as segregated Au nano-clusters, as observed in the TEM images
shown later.

\begin{figure}[!t]
\centering
\includegraphics[width=0.95\linewidth]{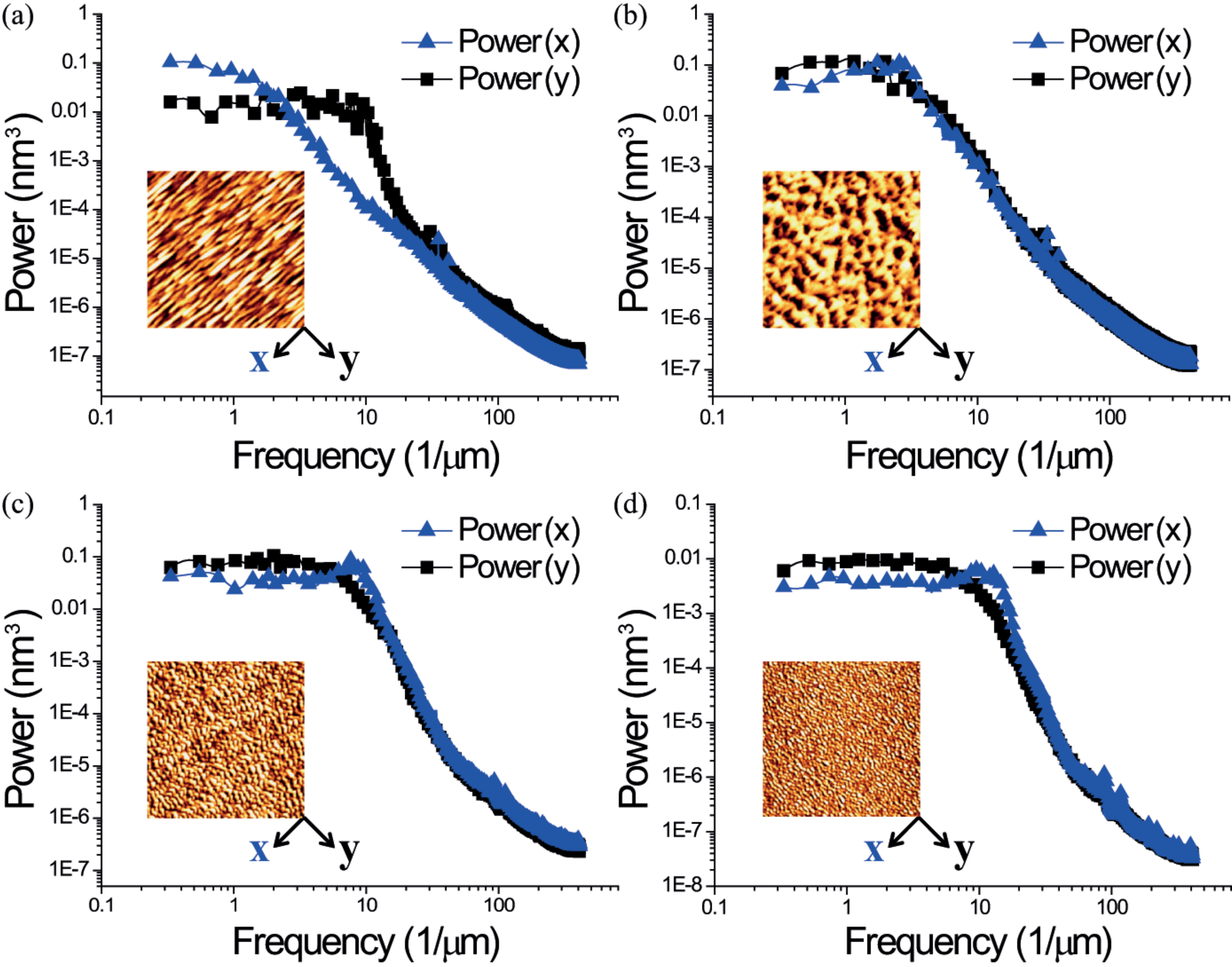}
\caption{1D power spectral densities of the surface
height at the four different sites considered in Fig.\
\ref{Patterns}, namely, (a) site 1 in the {\it relatively}
Au-rich region, (b) site 2 at the ripple reorientation
transition region, (c) site 3 in the Si-rich region, and (d)
site site 4 at the Si-richest region near the edge of the
sample. Triangles (squares) stand for the PSD data along the $x$($y$) direction.
}
\label{PSD}
\end{figure}

Figure~\ref{PSD}(a) shows the 1D power spectral densities (PSD) along the two
different directions on the substrate plane, namely,
the squared modulus of the FT of 1D cuts of the surface in Fig.\ \ref{Patterns}(a) along each
direction. From now on, $x$ corresponds to the projection
of the ion beam and $y$ is the perpendicular direction. In the figure
one can clearly identify a characteristic wave-vector value
along the $y$-axis, $k_y \simeq 10$ $\mu$m$^{-1}$, at which an
abrupt change takes place in the slope of the curve. This value
corresponds to the mean wavelength mode of the observed ripple
pattern. In contrast, there is no such feature along the
$x$-axis, parallel to the ion beam direction. Since the ripples
ridges are thus parallel to the latter, we term these {\it
parallel ripples}.

Figure\ \ref{Patterns}(b) shows the pattern observed 0.75 mm
away from site 1 and the Au source (site 2), as indicated
in the inset at the upper-right corner. As suggested by the 2D
autocorrelation and FT, patterning seems to be occurring along
the two substrate directions. Along the ion beam direction,
ripples still develop, with very short coherence lengths.
Perpendicular to the ion beam direction, shortened ripples
concatenate with neighboring ones to form an array of stripes.
The two 1D PSD curves along the $x$ and $y$ directions almost
coincide for this pattern, see Fig.\ \ref{PSD}(b), reflecting
the two-dimensional nature of the structure. Along the
$x$-direction, though, one can observe a broad peak, reflecting a
prominent ripple-like texture with a wave-vector running
along the ion beam direction. Such a quasi-two-dimensional
pattern is often observed during ripple reorientation
transitions.\cite{Metya:2013bq} In our system, the transition
properly takes place when one moves further away from the Au
source as detailed below. From now on, the region represented
by the site 2 is thus termed {\it transition region}.

Figure\ \ref{Patterns}(c) shows an image taken further away from the Au source, at site 3. A well-defined ripple pattern is
observed, but now the ripple ridges run perpendicular to the ion beam direction, so that ripple re-orientation has fully taken
place. We term these {\it perpendicular ripples}. Accordingly,
in Fig.\ \ref{PSD}(c) the PSD along the $x$-direction clearly
shows a sharp peak at a well-defined mean wavelength,
characteristic of the perpendicular ripple pattern.
Considering that, under the same sputtering condition, IBS
produces parallel ripples both for bulk\cite{Kim2009} Au(001)
and for the {\it relatively} Au-rich region in Fig.\
\ref{Patterns}(a),
the perpendicular ripple in Fig.\ \ref{Patterns}(c) and the
reorientation transition seem triggered by the reduced Au
concentration, sufficiently far away from the Au impurity
source.

\begin{figure} [!htbp]
\centering
\includegraphics[width=0.95\linewidth]{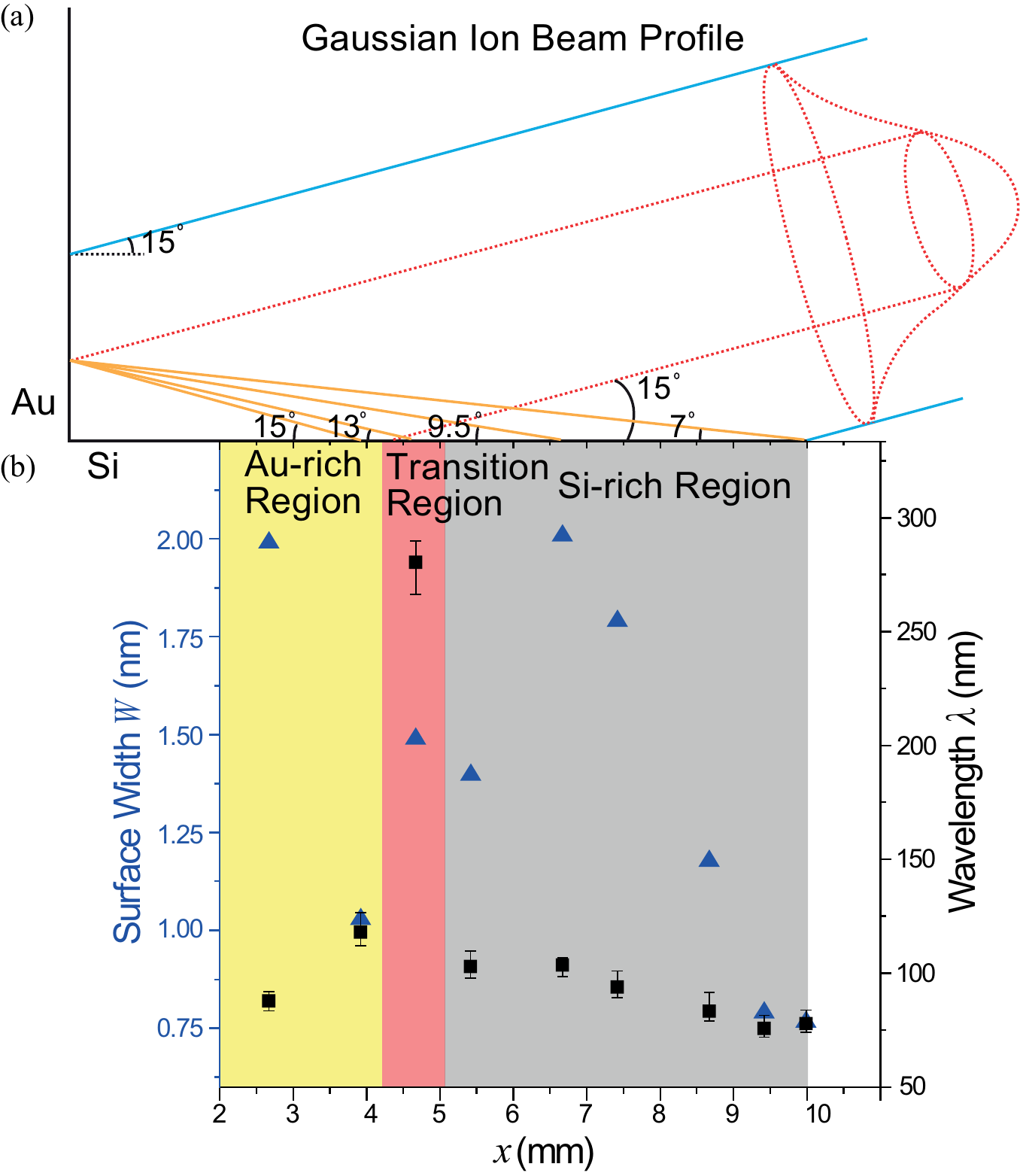}
\caption{(a) Sketch of the detailed experimental geometry.
Solid orange lines in (a) indicate the trajectories of
the recoiled or sputtered Au particles landing on the various sites
of the Si substrate. (b) The surface width ($W$, blue triangles) and the ripples wavelength
($\lambda$, black squares) as functions of the distance from the Au source.
Here, $x = 0$ indicates the position in contact with the Au
target, and $x = 10$ (mm) is the farthest position from the Au
source. Note, in the Au-rich (Si-rich) region ripples ridges are parallel (perpendicular)
to the projection of the ion beam.}
\label{Summary}
\end{figure}

Finally, Fig.\ \ref{Patterns}(d) shows an image at the farthest position from the Au source, site 4.
One can still observe a perpendicular ripple pattern, the
corresponding PSDs clearly showing a well-defined peak
along the $x$-direction [Fig.\ \ref{PSD}(d)], which corresponds to the
mean wavelength of these perpendicular ripples. The mean
wavelength and surface roughness of these ripples are smaller
than those at site 3, as summarized in Fig.\ \ref{Summary}(b).
Moving away from the Au source, and thus for a reduced Au flux,
the surface at site 4 behaves more like clean Si(100), which
shows very efficient healing kinetics leading to virtually flat
surfaces under the present sputtering condition, as shown in
Fig.\ \ref{Null_Pattern}.

The perpendicular ripples in both Figs.\ \ref{Patterns}(c) and
(d) are, however, wiggly; each ridge looks either sectioned
into small pieces or with a very small coherence length. This
suggests that the instability along the $y$ direction
perpendicular to the ripples still remains effective enough, so
as to induce sectioning of the ripple ridges. Moreover, each
piece has neither a uniform width nor a sinusoidal shape,
pointing to the significance of non-linear
effects.\cite{Kim2013,Munoz-Garcia2014}

Previous
experimental\cite{Hofsass2008,Macko2010,Macko2011,Zhang2011,Redondo-cubero2012,Hofsass2013apa,Engler2014}
and theoretical \cite{Bradley2012a,Bradley2013a} results indicate that the direction of the wave-vector of the ripple pattern
follows that of the impurity flux. Although these experimental
results are obtained for silicide-forming metallic impurities,
which is not the case in our system, the perpendicular ripple
patterns we observe at the sites 3 and 4 indeed seem consistent with
such results.
Figure\ \ref{Summary}(a) depicts the sputtering geometry and
also the recoil geometry in real scale, including sketches of
the directions of the Au impurity flux reaching the four sites.
In the previous experiments,\cite{Hofsass2008,Macko2010,Macko2011,Zhang2011,Redondo-cubero2012,Hofsass2013apa,Engler2014} the incidence of the ion
beam is near normal to the average surface orientation, or
its influence is isotropic. In our case, however, the ion beam
incidence is close to glancing.

In the {\it relatively} Au-rich region, the target behaves as a
pure Au surface in the erosive regime,\cite{Kim2009} for which
recrystallization is very efficient,\cite{Valbusa2002}
irradiation-induced material rearrangement or viscous flow is
negligible,\cite{Munoz-Garcia2014} and
the morphological instability seems to be of the erosive Bradley-Harper\cite{Bradley1988}
(BH) type.\cite{Chan2007}
This accounts for the parallel ripple orientation. In the
opposite limit in which Au impurities are scarce, purely erosive mechanisms are less effective in the Si-like surface (recall no
pattern forms in the absence of impurities), surface material
rearrangement or viscous flow being expected to be more
relevant.\cite{Munoz-Garcia2014} Still, the direction of the Au
flux may be influencing the orientation of the ripple pattern,
as frequently observed in other experiments. The 2D-like pattern in between the two, {\it relatively} Au-rich and Au-poor,
regions [Fig.\ \ref{Patterns}(b)] may result from the balance
of the two driving forces. We have performed SRIM
simulations\cite{Ziegler2010} (not shown) and obtained that the
sputtering yield of the recoiled Au impurities is negligible.
However, the recoiled Au atoms do transfer their momentum to the Si atoms and can displace them by close to 1 nm. This mass
displacement might drive the ripple orientation inducing
Carter-Vishnyakov-like (CV) surface-rearrangement
currents.\cite{Carter1996} Recall that, for high-incidence
angles, CV effects actually destabilize the surface and
contribute to ripple formation.

\begin{figure}[!t]
\centering
\includegraphics[width=0.85\linewidth]{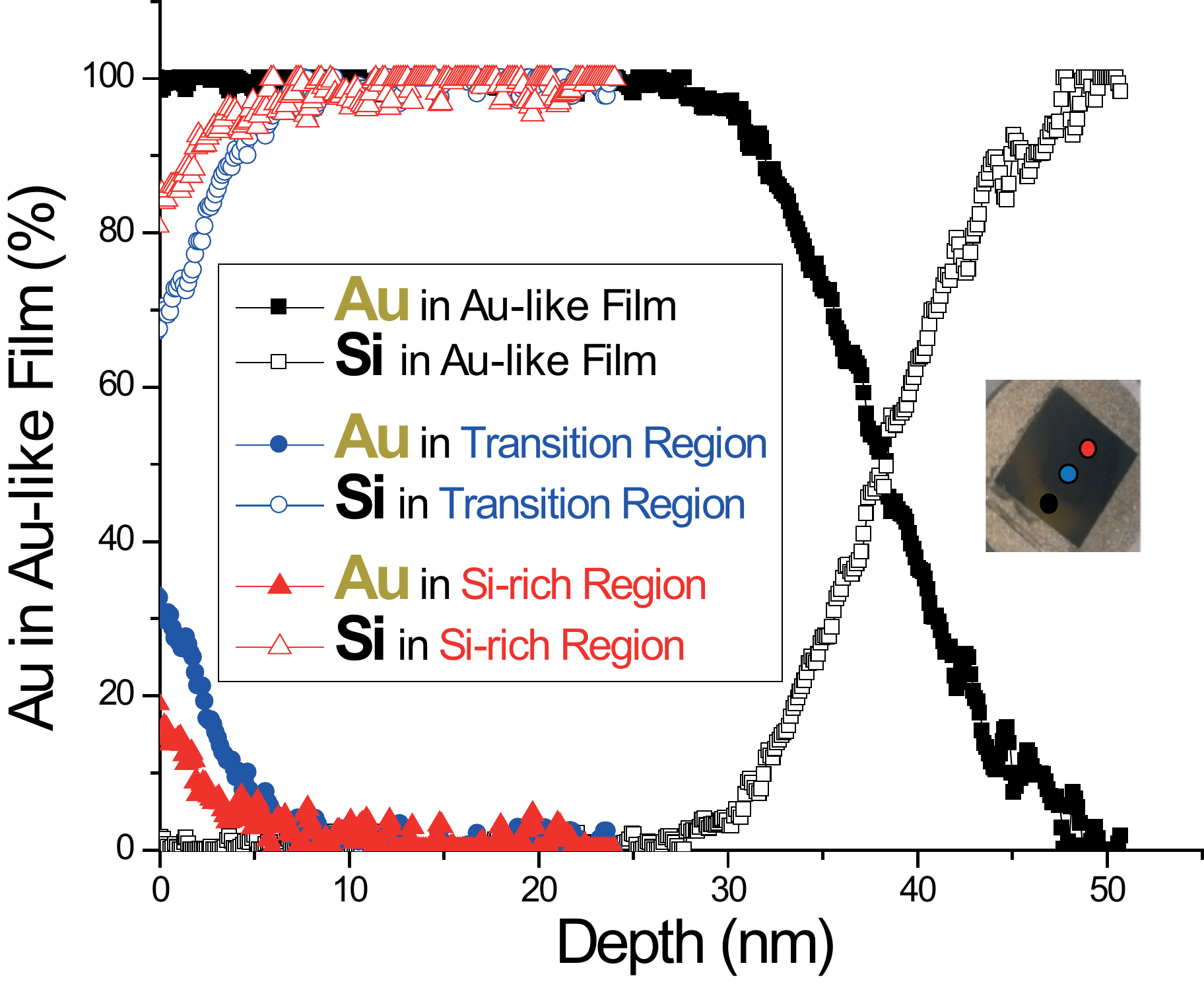}
\caption{Depth profiles of Au (MNN transition, 2,022 eV) and Si
(KLL transition, 1,621 eV) are the plots of AES
signals  while raster-sputtering 3$\times$3 mm$^2$ areas by
1 keV Ar$^{+}$. Note that the region named as 'Au-like film' is
sampled closer to the Au source than the site 1, as shown in the inset.}
\label{Depth_Profile}
\end{figure}

\begin{figure}[!t]
\centering
\includegraphics[width=0.95\linewidth]{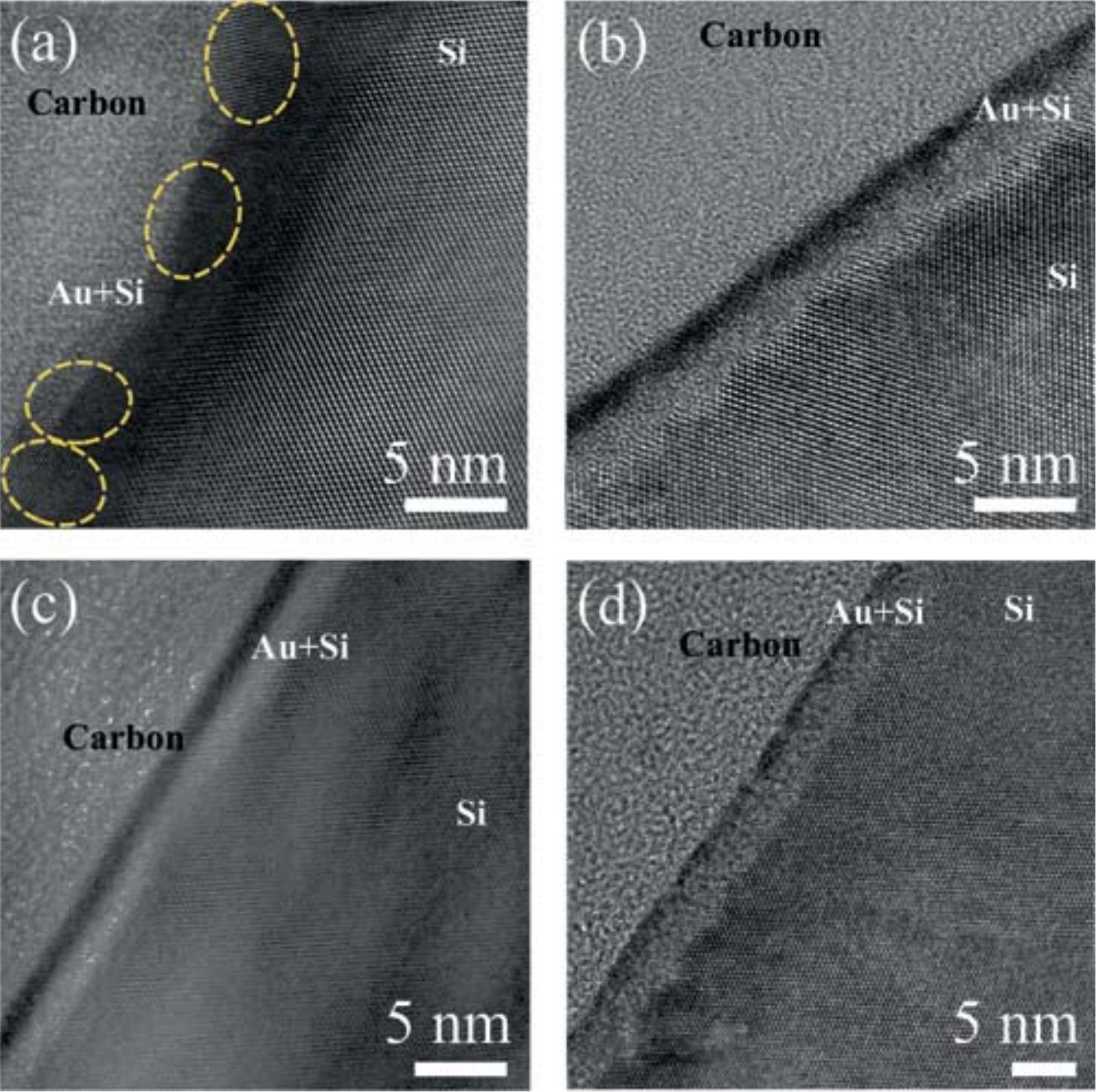}
\caption{HR-TEM images obtained from different regions of the Si target: (a) to (d), around the
sites 1 to 4 in Fig.~\ref{Patterns}, respectively. In (a), the yellow dotted circles
enclose the Au clusters. }
\label{HRTEM}
\end{figure}

\begin{figure} [!htbp]
\centering
\includegraphics[width=0.95\linewidth]{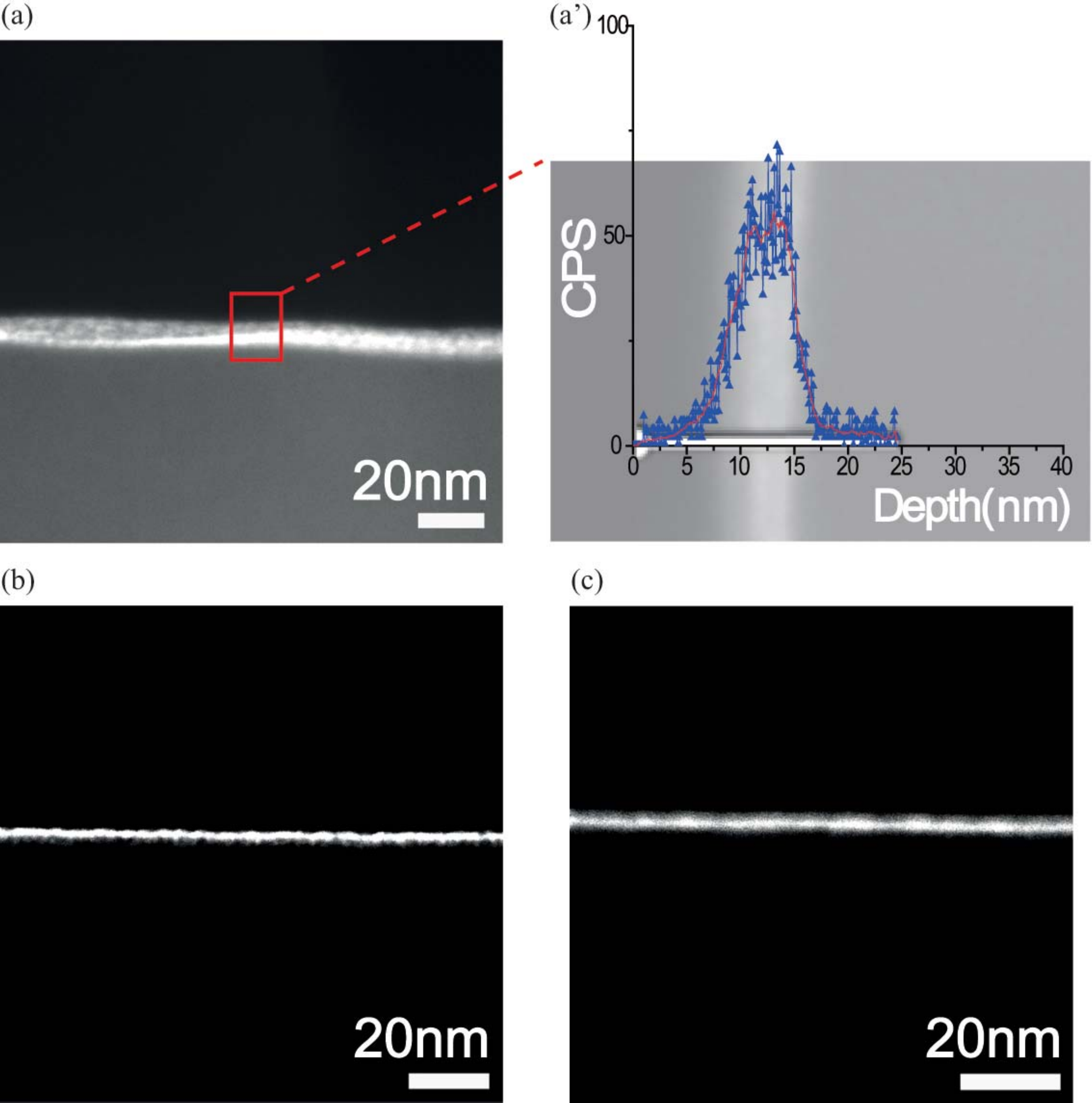}
\caption{ (a)-(c) HAADF-STEM images obtained from different regions around the sites 2, 3  and 4, respectively.  (a') EDX (Au(Lα1)) line profile of the boxed areas indicated in the HAADF-STEM image (a).
The EDX profile is displayed simultaneously with the HAADF-STEM image for comparison.}
\label{STEM}
\end{figure}

Figure~\ref{Depth_Profile} shows depth profiles of our samples
showing the atomic weight percent of Au and Si obtained by AES.
The measurement is made at three different locations: At the
vicinity of the Au source, in between sites 1 and 2, and close
to site 3, as shown in the inset of Fig.~\ref{Depth_Profile}. Since the depth profile is taken
over a 3 mm$\times$3 mm, raster-sputtered area, the spatial
resolution is limited by the same scale. From the profiles, we
can clearly see that, as expected, the Au content is higher for
locations which are closer to the Au source. The nominal atomic
weight percentages of Au at the surface decrease from 100\% to 30\%
(5.8\% of the atomic concentration), and then to 18\% (3.0\% of
the atomic concentration) as moving further away from the Au
source. The values are nominal, since they are calculated
by assuming that the concentration of Au is vertically uniform, or
 at least within the escape depth of the Au AES electrons. The
residual Au impurities are confined within a layer
which is approximately 5 nm deep, except for the Au-like film formed
closest to the Au target. This thickness is comparable to that
of the amorphized topmost surface layer that is revealed by
HRTEM and STEM (Figs.\ \ref{HRTEM} and \ref{STEM}). Note that
the amplitude of the nanostructures formed by IBS is less than 2 nm [cf.\ Fig.~\ref{Summary}(b)].
Thus, the nanostructures form solely in the Au containing
region, indicative of the active role of the Au impurities in
the pattern formation.

Fig.\ \ref{HRTEM}(a), the HRTEM image around site 1, shows
that the surface region is made of an amorphous Si layer embracing
Au nanoclusters. The yellow dotted  circles
indicate crystalline clusters displaying regularly spaced lines
oriented differently in different regions, but having distinctly different spacing from that of crystalline Si.
These clusters
seem to be formed of the co-deposited, but segregated, Au
impurities, due to the high Au concentration in this region. On
the other hand, their surroundings show no ordered features,
and are reminiscent of
an amorphous region formed of Si possibly with very disperse Au
impurities. For regions which are further away from the Au source [panels (b) through (d) in Fig.\ \ref{HRTEM}], no Au
nanoclusters in the amorphized layer can be detected by HRTEM,
the layer structure looking more homogeneous from this point of
view.

We have further characterized the chemical structure of our
samples through the EDX  in conjunction with High-angle
annular dark-field scanning transmission electron microscope
(HAADF-STEM).
The HAADF-STEM image in Fig.\ \ref{STEM}(a) shows
the interface with clear contrast in its intensity.
Fig.\ \ref{STEM}(a') shows the Au EDX signal along a
line normal to the interface in the boxed region in Fig.\ \ref{STEM}(a).
It clealy tells that the high intensity in Fig.\ \ref{STEM}(a) originates from the high Au concentration. Since Au does not form stable
silicides,\cite{Hofsass2013apa} we should expect phase
separation in the form of Au clusters.
Metastable clusters of Au silicide have been reported only at
elevated temperatures.\cite{Baumann1991}



Within the alloyed layer in Fig.\ \ref{STEM}(a), the Au impurity concentration looks higher around the ridge
than around valleys. High impurity concentration around ridges is commonly
observed for many silicide-forming
impurities.\cite{Hofsass2008,Macko2010,Macko2011,Zhang2011,Redondo-cubero2012,Hofsass2013apa,Engler2014}
In those systems, phase separation associated with silicide
formation is thought to be at the origin of the surface
nanopatterning. In principle, it is thus tempting to conclude
that phase separation of Au, and an inhomogeneous sputtering
yield distribution, also induces the observed pattern formation
in our case.

\begin{figure} [!tbp]
\centering
\includegraphics[width=0.95\linewidth]{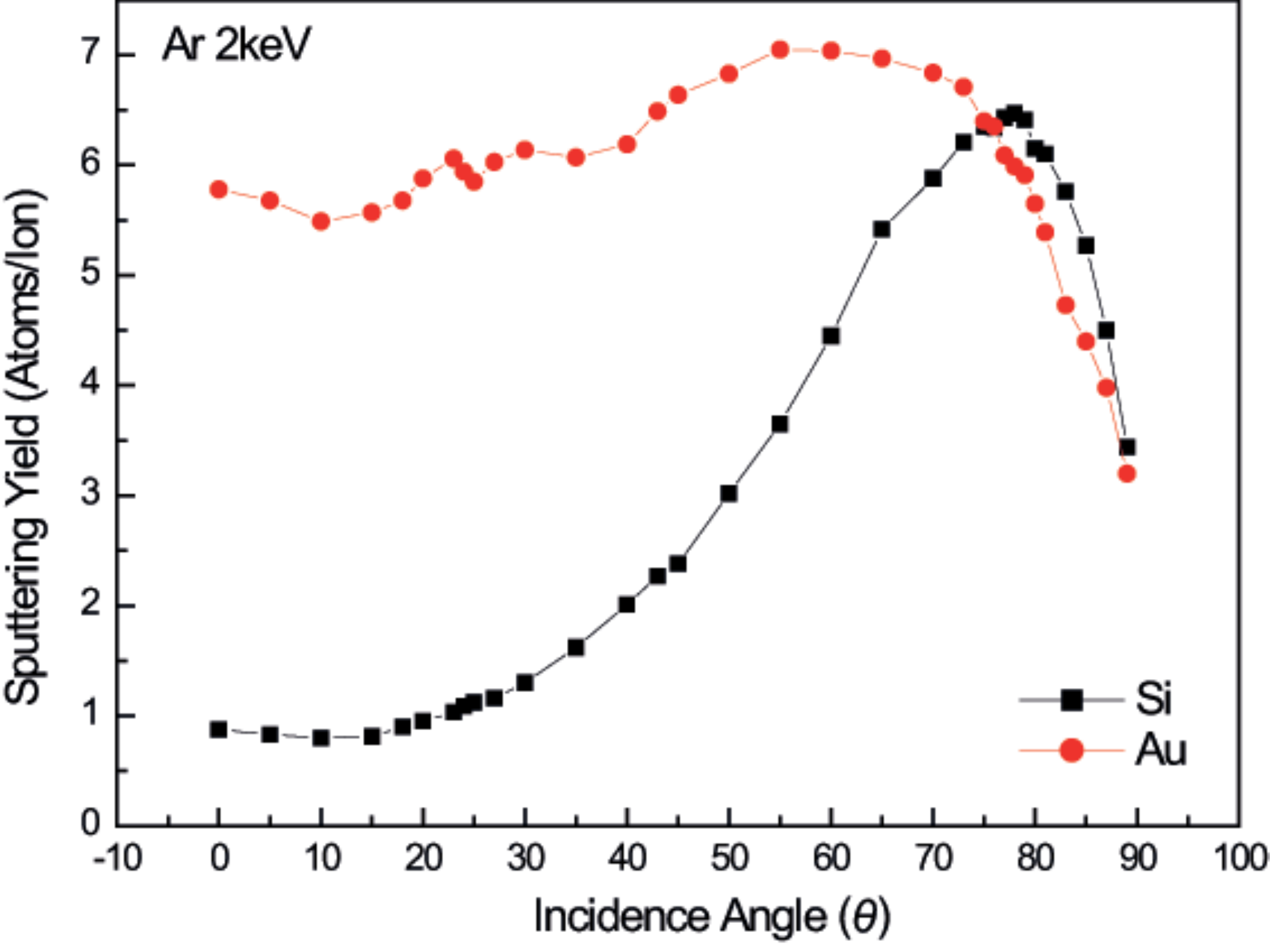}
\caption{Sputtering yields of Si (black squares) and Au (red bullets) targets vs incidence angle $\theta$, as obtained from SRIM simulations for 2 keV Ar$^+$ irradiation. Note the curves cross at the $\theta=75^{\circ}$ value employed in our experiments.}
\label{fig:srim}
\end{figure}

However, note that SRIM calculations\cite{Ziegler2010} under our present condition lead to virtually identical sputtering yields for Au and Si, see Fig.\ \ref{fig:srim}.
Hence, apparently phase
separation does not lead to an inhomogeneous sputtering yield
over the surface, and thus to pattern formation. Still, as
recently pointed out,\cite{Engler2014} rapid segregation of Au
under a constant external deposition flux makes the apparent
sputtering yield of phase-separated Au much smaller than that of Si,
since the external Au supply constantly replenishes the
sputtered Au. This effectively-lower Au sputtering yield, then,
might in turn lead to pattern formation. Within this scenario,
surface valleys erode faster due to the relatively low-Au
concentrations,
as compared with the ridge region.

Further away from the Au source, the Au flux is low and the
alloyed layer is thin as shown in Fig.\ \ref{STEM}(b)
and (c) consistently with the narrowed, (dark) Au containing layers
in Fig.\ \ref{HRTEM}(c) and (d), respectively.
 They do not reveal clear modulation of the Au concentration in the
substrate plane. Thus, at this moment we may not be able to contend
that the patterns observed in the {\it relatively} Au-poor region have
the same origin as in the  Au-rich region. Note, our failure to
observe a compositional pattern in the impurity-poor regions
may be due to the low Au concentration there. Combined with a low
spatial resolution of HAADF-STEM image,  this result may hamper detection of
small modulations of the impurity concentration, even if they
actually exist.

\section{Comparison with continuum models}
\label{sec:models}

To date, a number of continuum models are available in the
literature (some of which have been already mentioned), in
which surface nanopattern formation is described in the presence
of non-negligible impurity co-deposition, see an overview in
Ref.\ \onlinecite{Munoz-Garcia2014}. For instance, an early work for
so-called surfactant sputtering \cite{Kree2009} put forward a
coupled system of equations \cite{Shenoy2007,Cuerno2011}
for the dynamics of the surface height and the surfactant concentration,
although no predictions were provided for dependencies of the
pattern properties with experimental parameters. More recently,
related approaches have been pursued in greater detail,
considering the effects of various incidence conditions and
relaxation mechanisms. Thus, concurrent impingement of ions and
impurities has been considered,
\cite{Zhou2010,Bradley2011b,Bradley2012b} leading to e.g.\ an
analytical result \cite{Bradley2011b} on a minimal threshold
value which is required for the impurity concentration so that
patterns can appear. A generalization to oblique ion and
impurity incidences has then been performed,\cite{Bradley2012a}
with the result that an instability can arise purely as a result
of differential sputtering rates for the two impinging species,
in interplay with a phase shift between the concentration and
the height profiles.
More recently,\cite{Bradley2013a} silicide formation has been
explicitly incorporated to the models, with the conclusion that
it plays a decisive role, in agreement with many experimental
results as already discussed.

In this section we put the results of the experiments described
in Sections \ref{sec:exp} and \ref{sec:results} in the light of
these continuum models, with the aim to stress similarities
and differences between predictions and observations. Although
such a discussion will allow to somewhat rationalize the latter,
it will mostly suggest issues that should possibly be taken
into account for model improvement, to be able to account for the
experimental results.

Two very basic experimental observations can be taken as a first
basis for modeling: {\it (i)} Under the chosen irradiation
conditions, pure Si targets are morphologically stable, while Au
targets are not. {\it (ii)} There is a space gradient in the Au
impurity concentration within the Si target: this concentration
is maximum near the Au source and it decreases with distance to
the latter. Moreover, we can add the result from our SRIM
simulations shown in Fig.\ \ref{fig:srim} that, {\it (iii)} at the ion energy and incidence
angle considered, the sputtering yields of Au and Si take the
same values.

These observations can be readily implemented in the
phenomenological model originally put forward in Ref.\
\onlinecite{Bradley2011b} for ion-beam sputtering of a
monoelemental target of atomic species $B$ (silicon in our
experiment), under simultaneous co-deposition of impurities
of a different species $A$ (gold, in our case). Further important
assumptions which agree with our experimental results include
that the two atomic species are mutually inert, i.e.\ they form
no compound; and that a surface layer, of thickness $\Delta$,
forms on top of the irradiated target, within which there is
deposition of species $A$, both species being subject to
transport and sputtering effects. We should note that in this
model \cite{Bradley2011b} ions and impurities arrive under
normal incidence, in stark difference with the present
experiments. However, for the sake of simplicity we first
proceed by neglecting this fact. Likewise, we restrict ourselves
to a one-dimensional system. We will come back to these
assumptions later.

The model consists of the following two coupled
equations:\cite{Bradley2011b}
\begin{eqnarray}
\partial_t h(\mathbf{x},t) & = & - \Omega \left( F_A + F_B - F_d+ \nabla \cdot \mathbf{J}_A + \nabla \cdot \mathbf{J}_B\right) ,\label{eqh} \\
\Delta \partial_t c_s(\mathbf{x},t) & = & - \Omega \left( F_A -
F_d + \nabla \cdot \mathbf{J}_A \right) . \label{eqc}
\end{eqnarray}
Here, $F_d$ is the deposition flux of impurities, $c_s=c_A$
their space-time-dependent surface concentration (such that $c_B= 1-c_A$), $\Omega$ is the atomic volume of both species,
assumed to be equal for simplicity, $h$ is the height of the
irradiated target, and $F_i$ and $\mathbf{J}_i$ are,
respectively, the erosion flux and surface current of species
$i=A,B$. More specifically,
\begin{equation}
F_i = c_i \lambda_i (P_0 + \alpha_2 \nabla^2 h), \label{eq_f}
\end{equation}
where $P_0$ is a constant and $\lambda_i$ is the sputtering
yield of the $i$ species, such that for $\alpha_2>0$ erosion is
more efficient at surface troughs than at surface peaks, as in
the classic Bradley-Harper (BH) mechanism.\cite{Bradley1988} Finally, the surface current is
\begin{equation}
\mathbf{J}_i = -D_i \rho_s \nabla c_i + \frac{D_i \rho_s c_i
\Omega \gamma}{k_BT} \nabla \nabla^2 h - \mu_i c_i \nabla h,
\label{eq_j}
\end{equation}
where the first term on the rhs is Fickian diffusion ($D_i$ is
surface diffusivity and $\rho_s$ the areal density of mobile
surface atoms), the second one is Mullins' surface diffusion
($\gamma$ is surface tension and $T$ is temperature), and the
last one is a stabilizing Carter-Vishnyakov (CV)
term,\cite{Carter1996} in which $\mu_i >0$. For our experimental
system, for which the topmost surface layer can be thought of as an amorphous Si phase with Au impurities, CV terms can be
thought of as proxies of surface-confined viscous flow, recently shown to describe IBS patterning of clean Si
targets.\cite{Umbach2001,Castro2010,Cuerno2011,Castro2012a,Castro2012b,Norris2012,Moreno-Barrado2015,Moreno-Barrado2015b,Moreno-Barrado2015c}
In such a case, the coefficient in front of the third order
derivative term in $\mathbf{J}_B$ should more properly be
related with radiation-induced viscosity. Nevertheless, this
fact does not play a relevant role in the discussion to follow.
Similarly, note that pure Au targets recrystallize very
efficiently under our experimental conditions, so that $\mu_A$
is small as discussed above, at least in regions where the Au
density is high enough that recrystallization becomes feasible.

The analysis of model (\ref{eqh})-(\ref{eq_j}) performed in
Ref.\ \onlinecite{Bradley2011b} corresponds to the case of
immobile impurities, i.e., $D_A=\mu_A=0$. Moreover, a coordinate rescaling is performed \cite{Bradley2011b,Bradley2012b} by
constants which are inversely proportional to the sputtering
yield difference, $(\lambda_A - \lambda_B)$. But, because of the SRIM result from Fig.\ \ref{fig:srim}, we are interested in a system with the same sputtering yields for both species, for which this parameter combination is zero.

We have performed a linear stability analysis of the full model
(\ref{eqh})-(\ref{eq_j}) for precisely the case in which
$\lambda_A = \lambda_B \equiv \lambda$ and $D_A, \mu_A\neq 0$.
The result is that the model supports a flat solution $h=- v_0
t$ and $c_s = c_0$, characterized by a constant surface velocity $v_0=\Omega (\lambda P_0-F_d)$ and uniform impurity coverage
$c_0=F_d/(\lambda P_0)$. Moreover, under the standard
large-wavelength approximation (namely, for wave-vectors $k \ll
1$), the linear dispersion relation $\omega_k$ for periodic
perturbations of this flat solution, $h(x,t) =-v_0t+ u_*
e^{\omega_k t} e^{{\rm i} k x}$ and $c(x,t) = c_0 + \phi_*
e^{\omega_k t} e^{{\rm i} k x}$, where $|u_*|, |\phi_*| \ll
1$,\cite{Bradley2011b} reads
\begin{equation}
{\rm Re}(\omega_k^{+}) \simeq -C k^2 - G k^4 . \label{eq_wk}
\end{equation}
Naturally, both $G$ and $C$ depend on the parameters entering
model (\ref{eqh})-(\ref{eq_j}). For our case of interest, we
obtain
\begin{equation}
C = \Omega\left[ c_0 (\mu_A - \alpha_2 \lambda_A) + (1-c_0) (\mu_B- \alpha_2 \lambda_B) \right] \label{eq_C}
\end{equation}
and
\begin{equation}
G = \Omega \left[c_0 \Omega_A +(1-c_0)\Omega_B +c_0 (D_A-D_B)
\rho_s (\lambda \alpha_2-\mu_A)/(\lambda P_0) \right],
\label{eq_B}
\end{equation}
where $\Omega_i=\Omega {D_i \rho_s \gamma}/({k_BT})$.

The morphological stability of the surface is controlled by the
sign of the constant $C$ in Eq.\ \eqref{eq_C}, in a way that in
principle agrees with the simplest expectations: If the impurity concentration is very large ($c_0\simeq 1$), then $C \simeq
\Omega (\mu_A - \alpha_2 \lambda_A) < 0$ because gold is under a
pattern-forming condition (CV effects, $\mu_A$, are negligible
with respect to BH effects, $\alpha_2 \lambda_A$), thus the system displays pattern formation. Conversely, for very low impurity
concentration ($c_0\simeq 0$), then $C \simeq \Omega (\mu_B -
\alpha_2 \lambda_B) > 0$, which rules out unstable modes and
pattern formation, because Si is under such stable conditions
(CV contributions, $\mu_B$, dominate over BH effects, $\alpha_2
\lambda_B$). Actually, there is a threshold impurity
concentration, $c_0^* = (\mu_B - \alpha_2 \lambda_B)/(\mu_B -
\mu_A)$, such that patterns form only provided $c_0>c_0^*$.

Thus, model (\ref{eqh})-(\ref{eq_j}) can in principle
rationalize the basic experimental fact that Au impurities lead
to pattern formation in our Si targets. However, a number of
further considerations have to be made at this point:

\begin{itemize}
\item Although in the experimental system the reference surface
concentration of impurities $c_0 \propto F_d$ is {\it not} a
space-independent constant as assumed in the model, the
variation of the deposition flux across the experimental system
does occur in macroscopic scales which are much larger than the
ripple wavelength. On the other hand, in our experiments we
obtain ripple formation throughout the sample, while the model
leads to expecting flat surfaces for regions where $c_0 <
c_0^*$. This was checked in Fig.~\ref{Null_Pattern}, but perhaps
we did not employ large enough targets, with areas sufficiently
far from the Au source, featuring no surface pattern.

\item Whenever pattern formation occurs in the model topography,
a space modulation simultaneously takes place in the composition field. As discussed in Sec.\ \ref{sec:results}, in the
experiments such type of composition pattern is seen in the
relatively Au-rich regions only, being possibly beyond detection limit in the Au-poorer regions, for the experimental technique
employed.

\item The model suggests that, in regions where the impurity
concentration $c_0$ takes intermediate values, ripple formation
is analogous to what is seen for a pure Au system. However, this is not the case in the experiments either: in the former case
perpendicular ripples are observed, while parallel ripples form
on the relatively Au-rich regions which are very close to the
impurity source.
\end{itemize}

The lack of simultaneous patterning in the surface morphology
and composition suggests the need for a closer experimental
characterization of the space distribution of atomic species in
the amorphized surface layer that ensues. From the general point of view of reaction-diffusion type models,\cite{Cross2009} like
the present two-field model
(\ref{eqh})-(\ref{eq_j}),\cite{Munoz-Garcia2014} simultaneous
patterning of the two fields, i.e., height and impurity
concentration, is the rule rather than the exception.

On the other hand, indeed the assumption of normal incidence for both, ions and impurity atoms recoiling from the gold source, is an overly simplifying one. On the basis of the mentioned IBS
experiments on Si targets with concurrent silicide-forming
impurity co-deposition, a high relative angle between the ions
and impurities has been suggested to facilitate pattern
formation.\cite{Engler2014} Moreover, the results in Sect.\
\ref{sec:results} show a rotation of the ripple structure with
increasing Au concentration. Thus, a two-dimensional
generalization of model (\ref{eqh})-(\ref{eq_j}) is required,
for arbitrary ion and impurity incidence angles. This will lead
in particular to an anisotropic version of the linear dispersion relation (\ref{eq_wk})-(\ref{eq_B}). However, it is not obvious
that this can improve the results on the ripple orientation as a function of impurity concentration $c_0$. This is because,
experimentally, the $y$ axis is the only unstable direction for
Au, while both directions are stable for Si. Therefore, a convex linear combination of the (2D anisotropic generalizations of
the) Au and Si 2D linear dispersion relations, such as Eq.\
\eqref{eq_C} is, cannot possibly yield an unstable $x$
direction, as would be needed to explain the ripple orientation
for intermediate impurity concentrations.

In the search for alternative models to account for the present
observations, we note that, although formulated for
(impurity-free) IBS of binary systems, a model has been put
forward \cite{Norris2013b} in which (ion-assisted)
phase-separation can control the nano structuring process. In
this work it is concluded that pattern formation will occur
only if phase separation in the amorphized layer is fast enough so
that it completes before the layer is sputtered away. Moreover,
the morphological transition in such a case leads to highly
ordered patterns. The large degree of disorder in the
perpendicular ripple structures we observe, and the relative
homogeneity of the amorphous layer in the corresponding regions,
both contrast with these theoretical results.

Seeking for further theoretical descriptions, recall that, for
pure Si targets, ripple formation has been recently accounted
for on the basis of viscous flow of the topmost amorphized
layer.\cite{Castro2010,Cuerno2011,Castro2012a,Castro2012b,Norris2012,Moreno-Barrado2015,Moreno-Barrado2015b,Moreno-Barrado2015c}In
this approach, a crucial effect of ion irradiation is
inducing residual stress in the amorphous layer, which is
relaxed via solid flow. The characteristics of the surface
dynamics are contingent upon the properties of the
non-homogeneous stress distribution that builds up within the
layer,\cite{Moreno-Barrado2015,Moreno-Barrado2015b} which in
particular controls the value of the critical incidence
$\theta_c$ angle for the ions, above which perpendicular ripples form.
In our present experiment, ripple formation does not occur on uncontaminated Si, namely $\theta_c \gtrsim 90^{\circ}$,
which we expect to originate in the properties of the stress
distribution under the corresponding conditions. Recently, a
similar result has been accounted for \cite{Hofsass2015} on the
basis of material redistribution using a Monte Carlo-based
crater function approach, which we believe can constitute an
equivalent, microscopic-based description of viscous flow.
Under this scenario, perpendicular ripple formation
in our experiments for intermediate impurity concentration might occur due to the
influence of Au impurities in the stress distribution. Indeed,
recall the results of our SRIM simulations that recoiled Au
atoms transfer momentum to the Si atoms and can displace them by close to 1 nm.
Such a displacement may reflect into a modified
stress distribution, to such an extent that it may be
responsible for the formation of perpendicular ripples. In the
relatively Au-rich regions, due to efficient recrystallization,
the high gold coverage would override viscous flow and lead to
parallel ripple formation, akin to a pure Au target under the
same sputtering conditions.


\section{Discussion}
\label{sec:disc}

%
%
%

Our experimental results show that, under conditions for which
pure Si targets do not become structured under IBS, Au
impurities can induce surface pattern formation, in a way that
is correlated with the impurity concentration. This is in spite
of the fact that, as discussed in detail
elsewhere,\cite{Hofsass2013apa} no silicide is expected to form
for our combination or materials. Already at the low Au
concentration values reached far ($\gtrsim 10$ mm) from the
impurity source, a perpendicular ripple pattern forms, which is
strongly disordered and has small wavelength and amplitude. The
orientation of these ripples conforms to expectations based on
silicide-forming impurities,\cite{Engler2014} that a large
relative angle between ions and recoiling impurity
atoms, $\alpha$, enhances pattern formation. In our case, such an angle
value is large indeed, $\alpha \simeq 150^{\circ}$. This fact
may account for the disagreement between some of our
experimental observations and currently available models, usually
studied for small $\alpha$.

Closer to the gold plate, for moderate but increasing impurity concentration values, the
wavelength of the perpendicular ripples also
increases. However, in this range of Au concentration values we
have not been able to detect a direct correlation between the
behaviors of the topography and the composition, i.e.,
we do not find a space modulation of Au concentration field. If there is
any, it remains beyond detection limit for the STEM measurements
reported in Sec.\ \ref{sec:results}. This fact calls for a more
detailed experimental assessment of the impurity concentration,
with respect to both, the substrate coordinates (composition
modulation) and the depth along the thickness of the amorphized
layer.

Actually, co-deposition of Au impurities has been already
attempted previously\cite{Hofsass2013apa,Engler2014} in order to surface pattern
Si through IBS, to no effect. The experimental conditions
employed in these works are nevertheless different from ours,
the most notable difference is our large relative angle
between ions and the recoiling Au atoms. Engler {\it et al}.\
already \cite{Engler2014} suggested experimental setups with
large $\alpha$ to maximize the coupling between height
fluctuations and the concentration modulation via shadowing
effects. A second major feature of our setup is the large
ion incidence angle, $\theta$, employed. In our case $\theta=75^\circ$. This also stems from
an observation by Engler {\it et al.} They observed that large incidence angles drive Si marginally stable, already without Au co-deposition.
These might have also triggered our system into the unstable state, giving
the patterns we observe. As a third difference with previous works, we have used Ar$^+$
ions, which are lighter than previous choices, such as Xe$^+$ or
Kr$^+$. Note that the use of light elements as projectiles does
not necessarily enhance pattern formation in the absence of
impurity co-deposition. Thus, under our sputtering condition
Ar$^+$ does not induce ripple formation on clean Si, while
Kr$^+$ does.\cite{Engler2014} Indeed, the ion/target mass ratio
is known to play a non-trivial role
in the ion-induced stress distribution and thus in the
patterning properties of
Si.\cite{Moreno-Barrado2015,Moreno-Barrado2015c}
Increased momentum transfer from the use of
heavy projectiles is supposed to drive the Si substrate into the
unstable state. In short, our experimental combination of relatively high
$\alpha$ and $\theta$ values seems to cooperatively destabilize
the Si surface under IBS for relatively low impurity
concentrations. Then, the space modulation of the latter,
coupled to height fluctuations, might have led to the observed
perpendicular ripple patterns. This picture needs to be taken
with caution, since precisely in the Au-poor regions we cannot
detect the modulation of the Au concentration profile.

For locations which are sufficiently close to the Au source,
impurities are able to form clusters, as seen in our HR TEM
images, which appear to be dispersed within an amorphous Si
matrix. Morphologically, this seems sufficient for the surface
to display parallel ripples, very much akin to those obtained on
pure Au films. The concentration of Au which is required for
this behavior has a moderate value near $6\%$. Close to it, the
morphological behavior corresponds to a transition between the
topographies observed for relatively low and relatively high Au
concentrations. Thus, somehow a superposition of the two ripple
structures is observed, characterized (recall the PSD curves in
Fig.\ \ref{PSD}) by an increase of the ripple wavelength in the
$x$ direction until disappearance of the characteristic scale
along this direction for increasing Au concentration. This occurs together
with simultaneous appearance of a characteristic wavelength
along the $y$ direction. The behavior of the PSD curves seems
reminiscent\cite{Cross2009} of a morphological Type II
transition along the $x$ direction and a Type I transition along the $y$ direction: indeed,\cite{Cross2009} in a Type II
transition the characteristic wavelength diverges when
approaching the transition point from the pattern-forming side.
This is actually the experimentally observed behavior for pure
sapphire\cite{Zhou2007} or Si targets in the absence of
impurities at low
energies.\cite{Madi2008,Madi2009,Madi2011,Madi2012,Castro2012b,Moreno-Barrado2015}
In contrast, a Type I morphological transition features the
sudden appearance of a characteristic wavelength at the
corresponding transition point.\cite{Cross2009} To date, there
seem to be no experimental observations of Type I transitions in
the context of IBS surface nanopatterning.\cite{Madi2012} Note
an alternative scenario is actually feasible for our
experimental results, as already discussed above: The coherence
length along the $y$ axis of the low-Au-concentration ripples
may be indicating that the corresponding wave-vector corresponds
to an unstable Fourier mode of the height, which is not the most dominant mode
for such Au concentration values, but becomes so
at a sufficiently high value of the impurity concentration.
Hence, systematic experimental confirmation of the type of
morphological transition that is actually taking place seems
required. In particular, this may provide an invaluable hint to
the theoretical modeling, as the transition type usually
constrains basic structural aspects of potential theoretical
descriptions.

On general grounds, the lack of compound formation and the fact
that ripples are actually induced on the Si target ---already
for very small impurity concentration values under otherwise
non-pattern forming conditions--- leads one naturally to
contemplate the relevance of viscous flow as the mechanism
controlling the surface dynamics. Indeed, it has been recently
shown\cite{Moreno-Barrado2015,Moreno-Barrado2015c}
for clean Si targets that modifications in the space distribution of
irradiation-induced stress, e.g.\ by changing the ion/target combination,
can alter the morphological stability of the surface. In the
present experiments, changes in the impurity concentration are
also shown to have a similar impact, hence it is natural to
ponder whether they correlate with analogous modifications in
the stress distribution which could account to the observed
ripple properties. Microscopic, e.g.\ Molecular Dynamics,
simulations can be naturally expected to provide insights into
this issue.

\section{Summary and Conclusion}
\label{sec:concl}

We find that Au co-deposition can catalyze pattern formation on
Si(001) during IBS under a sputter condition producing no
patterns on clean Si(001). With the increase of Au flux
closer to the Au co-sputtered target, the ripple pattern on Si
changes its $k$-vector direction, parallel to perpendicular
to the surface-projected ion beam direction. Au does not form stable
silicides, so that the present observation is at variance with
the prevailing notion that silicide formation is a
prerequisite for impurity-induced pattern formation.

Within an existing minimal model coupling the dynamics of
the height and concentration fields, in which no silicide
formation takes place, the instability can still develop with
the increase of Au concentration. The present work, thus, provides
a first example demonstrating such a novel mechanism for
impurity-induced pattern formation without invoking silicide formation.
Improvement of the theoretical model is still required in order to reflect the
real experimental situation and elucidate many unattended observations, such as the ripple
reorientation transition.

As predicted by the model, we observe a modulation of the Au concentration in the patterns
formed with a relatively high impurity flux. Replenishing the Au coverage by the continuous influx of recoiled impurity atoms from the gold target reduces the effective sputtering
yield of gold near the Au segregated crest, thus the resulting inhomogeneity of the sputtering yield can promote pattern formation. For the relatively Au-poor region, such a spatial
inhomogeneity is not discernible for the impurity coverage, so that the generalization of the picture awaits further elaborate experimental investigation.

\section{Acknowledgments}

We would like to thank L. V\'azquez for discussions. This work
was supported by NRF (Korea) Grants Nos.\ 2014K2A1A2048433 and
2013R1A2000245 and by MINECO (Spain) Grants Nos.\ FIS2012-32349
and FIS2012-38866-C05-01.


\begin{thebibliography}{47}
\expandafter\ifx\csname natexlab\endcsname\relax\def\natexlab#1{#1}\fi
\expandafter\ifx\csname bibnamefont\endcsname\relax
  \def\bibnamefont#1{#1}\fi
\expandafter\ifx\csname bibfnamefont\endcsname\relax
  \def\bibfnamefont#1{#1}\fi
\expandafter\ifx\csname citenamefont\endcsname\relax
  \def\citenamefont#1{#1}\fi
\expandafter\ifx\csname url\endcsname\relax
  \def\url#1{\texttt{#1}}\fi
\expandafter\ifx\csname urlprefix\endcsname\relax\def\urlprefix{URL }\fi
\providecommand{\bibinfo}[2]{#2}
\providecommand{\eprint}[2][]{\url{#2}}

\bibitem[{\citenamefont{Mu{\~n}oz-Garc{\'\i}a
  et~al.}(2009)\citenamefont{Mu{\~n}oz-Garc{\'\i}a, V{\'a}zquez, Cuerno,
  S{\'a}nchez-Garc{\'\i}a, Castro, and Gago}}]{Munoz-Garcia2009book}
\bibinfo{author}{\bibfnamefont{J.}~\bibnamefont{Mu{\~n}oz-Garc{\'\i}a}},
  \bibinfo{author}{\bibfnamefont{L.}~\bibnamefont{V{\'a}zquez}},
  \bibinfo{author}{\bibfnamefont{R.}~\bibnamefont{Cuerno}},
  \bibinfo{author}{\bibfnamefont{J.~A.} \bibnamefont{S{\'a}nchez-Garc{\'\i}a}},
  \bibinfo{author}{\bibfnamefont{M.}~\bibnamefont{Castro}}, \bibnamefont{and}
  \bibinfo{author}{\bibfnamefont{R.}~\bibnamefont{Gago}}, in
  \emph{\bibinfo{booktitle}{Toward Functional Nanomaterials}}, edited by
  \bibinfo{editor}{\bibfnamefont{Z.~M.} \bibnamefont{Wang}}
  (\bibinfo{publisher}{Springer}, \bibinfo{year}{2009}), pp.
  \bibinfo{pages}{323--398}.

\bibitem[{\citenamefont{Facsko et~al.}(1999)\citenamefont{Facsko, Dekorsy,
  Koerdt, Trappe, Kurz, Vogt, and Hartnagel}}]{Facsko1999}
\bibinfo{author}{\bibfnamefont{S.}~\bibnamefont{Facsko}},
  \bibinfo{author}{\bibfnamefont{T.}~\bibnamefont{Dekorsy}},
  \bibinfo{author}{\bibfnamefont{C.}~\bibnamefont{Koerdt}},
  \bibinfo{author}{\bibfnamefont{C.}~\bibnamefont{Trappe}},
  \bibinfo{author}{\bibfnamefont{H.}~\bibnamefont{Kurz}},
  \bibinfo{author}{\bibfnamefont{A.}~\bibnamefont{Vogt}}, \bibnamefont{and}
  \bibinfo{author}{\bibfnamefont{H.~L.} \bibnamefont{Hartnagel}},
  \bibinfo{journal}{Science} \textbf{\bibinfo{volume}{285}},
  \bibinfo{pages}{1551} (\bibinfo{year}{1999}).

\bibitem[{\citenamefont{Mu{\~n}oz-Garc{\'\i}a
  et~al.}(2014)\citenamefont{Mu{\~n}oz-Garc{\'\i}a, V\'azquez, Castro, Gago,
  Redondo-Cubero, Moreno-Barrado, and Cuerno}}]{Munoz-Garcia2014}
\bibinfo{author}{\bibfnamefont{J.}~\bibnamefont{Mu{\~n}oz-Garc{\'\i}a}},
  \bibinfo{author}{\bibfnamefont{L.}~\bibnamefont{V\'azquez}},
  \bibinfo{author}{\bibfnamefont{M.}~\bibnamefont{Castro}},
  \bibinfo{author}{\bibfnamefont{R.}~\bibnamefont{Gago}},
  \bibinfo{author}{\bibfnamefont{A.}~\bibnamefont{Redondo-Cubero}},
  \bibinfo{author}{\bibfnamefont{A.}~\bibnamefont{Moreno-Barrado}},
  \bibnamefont{and} \bibinfo{author}{\bibfnamefont{R.}~\bibnamefont{Cuerno}},
  \bibinfo{journal}{Materials Science and Engineering: R: Reports}
  \textbf{\bibinfo{volume}{86}}, \bibinfo{pages}{1} (\bibinfo{year}{2014}).

\bibitem[{\citenamefont{Gago et~al.}(2001)\citenamefont{Gago, V\'{a}zquez,
  Cuerno, Varela, Ballesteros, and Albella}}]{Gago2001}
\bibinfo{author}{\bibfnamefont{R.}~\bibnamefont{Gago}},
  \bibinfo{author}{\bibfnamefont{L.}~\bibnamefont{V\'{a}zquez}},
  \bibinfo{author}{\bibfnamefont{R.}~\bibnamefont{Cuerno}},
  \bibinfo{author}{\bibfnamefont{M.}~\bibnamefont{Varela}},
  \bibinfo{author}{\bibfnamefont{C.}~\bibnamefont{Ballesteros}},
  \bibnamefont{and} \bibinfo{author}{\bibfnamefont{J.~M.}
  \bibnamefont{Albella}}, \bibinfo{journal}{Applied Physics Letters}
  \textbf{\bibinfo{volume}{78}}, \bibinfo{pages}{3316} (\bibinfo{year}{2001}).

\bibitem[{\citenamefont{Ozaydin et~al.}(2005)\citenamefont{Ozaydin, \"Ozcan,
  Wang, Ludwig, Zhou, Headrick, and Siddons}}]{Ozaydin2005}
\bibinfo{author}{\bibfnamefont{G.}~\bibnamefont{Ozaydin}},
  \bibinfo{author}{\bibfnamefont{A.~S.} \bibnamefont{\"Ozcan}},
  \bibinfo{author}{\bibfnamefont{Y.}~\bibnamefont{Wang}},
  \bibinfo{author}{\bibfnamefont{F.}~\bibnamefont{Ludwig}},
  \bibinfo{author}{\bibfnamefont{H.}~\bibnamefont{Zhou}},
  \bibinfo{author}{\bibfnamefont{R.~L.} \bibnamefont{Headrick}},
  \bibnamefont{and} \bibinfo{author}{\bibfnamefont{D.~P.}
  \bibnamefont{Siddons}}, \bibinfo{journal}{Applied Physics Letters}
  \textbf{\bibinfo{volume}{87}}, \bibinfo{pages}{163104}
  (\bibinfo{year}{2005}).

\bibitem[{\citenamefont{Zhou et~al.}(2011)\citenamefont{Zhou, Facsko, Lu, and
  Moller}}]{Zhou2011}
\bibinfo{author}{\bibfnamefont{J.}~\bibnamefont{Zhou}},
  \bibinfo{author}{\bibfnamefont{S.}~\bibnamefont{Facsko}},
  \bibinfo{author}{\bibfnamefont{M.}~\bibnamefont{Lu}}, \bibnamefont{and}
  \bibinfo{author}{\bibfnamefont{W.}~\bibnamefont{Moller}},
  \bibinfo{journal}{Journal of Applied Physics} \textbf{\bibinfo{volume}{109}},
  \bibinfo{pages}{104315} (\bibinfo{year}{2011}).

\bibitem[{\citenamefont{Madi et~al.}(2011)\citenamefont{Madi, Anzenberg,
  Ludwig~Jr, and Aziz}}]{Madi2011}
\bibinfo{author}{\bibfnamefont{C.}~\bibnamefont{Madi}},
  \bibinfo{author}{\bibfnamefont{E.}~\bibnamefont{Anzenberg}},
  \bibinfo{author}{\bibfnamefont{K.}~\bibnamefont{Ludwig~Jr}},
  \bibnamefont{and} \bibinfo{author}{\bibfnamefont{M.}~\bibnamefont{Aziz}},
  \bibinfo{journal}{Physical Review Letters} \textbf{\bibinfo{volume}{106}},
  \bibinfo{pages}{66101} (\bibinfo{year}{2011}).

\bibitem[{\citenamefont{Hofs{\"a}ss and Zhang}(2008)}]{Hofsass2008}
\bibinfo{author}{\bibfnamefont{H.}~\bibnamefont{Hofs{\"a}ss}} \bibnamefont{and}
  \bibinfo{author}{\bibfnamefont{K.}~\bibnamefont{Zhang}},
  \bibinfo{journal}{Applied Physics A} \textbf{\bibinfo{volume}{92}},
  \bibinfo{pages}{517} (\bibinfo{year}{2008}).

\bibitem[{\citenamefont{Macko et~al.}(2010)\citenamefont{Macko, Frost, Ziberi,
  F\"{o}rster, and Michely}}]{Macko2010}
\bibinfo{author}{\bibfnamefont{S.}~\bibnamefont{Macko}},
  \bibinfo{author}{\bibfnamefont{F.}~\bibnamefont{Frost}},
  \bibinfo{author}{\bibfnamefont{B.}~\bibnamefont{Ziberi}},
  \bibinfo{author}{\bibfnamefont{D.~F.} \bibnamefont{F\"{o}rster}},
  \bibnamefont{and} \bibinfo{author}{\bibfnamefont{T.}~\bibnamefont{Michely}},
  \bibinfo{journal}{Nanotechnology} \textbf{\bibinfo{volume}{21}},
  \bibinfo{pages}{085301} (\bibinfo{year}{2010}).

\bibitem[{\citenamefont{Zhang et~al.}(2011)\citenamefont{Zhang, Br{\"o}tzmann,
  and Hofs{\"a}ss}}]{Zhang2011}
\bibinfo{author}{\bibfnamefont{K.}~\bibnamefont{Zhang}},
  \bibinfo{author}{\bibfnamefont{M.}~\bibnamefont{Br{\"o}tzmann}},
  \bibnamefont{and}
  \bibinfo{author}{\bibfnamefont{H.}~\bibnamefont{Hofs{\"a}ss}},
  \bibinfo{journal}{New Journal of Physics} \textbf{\bibinfo{volume}{13}},
  \bibinfo{pages}{013033} (\bibinfo{year}{2011}).

\bibitem[{\citenamefont{Macko et~al.}(2011)\citenamefont{Macko, Frost, Engler,
  Hirsch, H{\"o}che, Grenzer, and Michely}}]{Macko2011}
\bibinfo{author}{\bibfnamefont{S.}~\bibnamefont{Macko}},
  \bibinfo{author}{\bibfnamefont{F.}~\bibnamefont{Frost}},
  \bibinfo{author}{\bibfnamefont{M.}~\bibnamefont{Engler}},
  \bibinfo{author}{\bibfnamefont{D.}~\bibnamefont{Hirsch}},
  \bibinfo{author}{\bibfnamefont{T.}~\bibnamefont{H{\"o}che}},
  \bibinfo{author}{\bibfnamefont{J.}~\bibnamefont{Grenzer}}, \bibnamefont{and}
  \bibinfo{author}{\bibfnamefont{T.}~\bibnamefont{Michely}},
  \bibinfo{journal}{New Journal of Physics} \textbf{\bibinfo{volume}{13}},
  \bibinfo{pages}{073017} (\bibinfo{year}{2011}).

\bibitem[{\citenamefont{Hofs{\"a}ss et~al.}(2013)\citenamefont{Hofs{\"a}ss,
  Bobes, and Zhang}}]{Hofsass2013}
\bibinfo{author}{\bibfnamefont{H.}~\bibnamefont{Hofs{\"a}ss}},
  \bibinfo{author}{\bibfnamefont{O.}~\bibnamefont{Bobes}}, \bibnamefont{and}
  \bibinfo{author}{\bibfnamefont{K.}~\bibnamefont{Zhang}},
  \bibinfo{journal}{AIP Conference Proceedings}
  \textbf{\bibinfo{volume}{1525}}, \bibinfo{pages}{386} (\bibinfo{year}{2013}).

\bibitem[{\citenamefont{Engler et~al.}(2014)\citenamefont{Engler, Frost,
  M{\"u}ller, Macko, Will, Feder, Spermann, H{\"u}bner, Facsko, and
  Michely}}]{Engler2014}
\bibinfo{author}{\bibfnamefont{M.}~\bibnamefont{Engler}},
  \bibinfo{author}{\bibfnamefont{F.}~\bibnamefont{Frost}},
  \bibinfo{author}{\bibfnamefont{S.}~\bibnamefont{M{\"u}ller}},
  \bibinfo{author}{\bibfnamefont{S.}~\bibnamefont{Macko}},
  \bibinfo{author}{\bibfnamefont{M.}~\bibnamefont{Will}},
  \bibinfo{author}{\bibfnamefont{R.}~\bibnamefont{Feder}},
  \bibinfo{author}{\bibfnamefont{D.}~\bibnamefont{Spermann}},
  \bibinfo{author}{\bibfnamefont{R.}~\bibnamefont{H{\"u}bner}},
  \bibinfo{author}{\bibfnamefont{S.}~\bibnamefont{Facsko}}, \bibnamefont{and}
  \bibinfo{author}{\bibfnamefont{T.}~\bibnamefont{Michely}},
  \bibinfo{journal}{Nanotechnolgy} \textbf{\bibinfo{volume}{25}},
  \bibinfo{pages}{115303} (\bibinfo{year}{2014}).

\bibitem[{\citenamefont{Bradley}(2013)}]{Bradley2013a}
\bibinfo{author}{\bibfnamefont{R.~M.} \bibnamefont{Bradley}},
  \bibinfo{journal}{Physical Review B} \textbf{\bibinfo{volume}{87}},
  \bibinfo{pages}{205408} (\bibinfo{year}{2013}).

\bibitem[{\citenamefont{Zhou and Lu}(2010)}]{Zhou2010}
\bibinfo{author}{\bibfnamefont{J.}~\bibnamefont{Zhou}} \bibnamefont{and}
  \bibinfo{author}{\bibfnamefont{M.}~\bibnamefont{Lu}},
  \bibinfo{journal}{Physical Review B} \textbf{\bibinfo{volume}{82}},
  \bibinfo{pages}{125404} (\bibinfo{year}{2010}).

\bibitem[{\citenamefont{Bradley}(2011)}]{Bradley2011b}
\bibinfo{author}{\bibfnamefont{R.~M.} \bibnamefont{Bradley}},
  \bibinfo{journal}{Physical Review B} \textbf{\bibinfo{volume}{83}},
  \bibinfo{pages}{195410} (\bibinfo{year}{2011}).

\bibitem[{\citenamefont{Bradley and Shipman}(2012)}]{Bradley2012b}
\bibinfo{author}{\bibfnamefont{R.~M.} \bibnamefont{Bradley}} \bibnamefont{and}
  \bibinfo{author}{\bibfnamefont{P.~D.} \bibnamefont{Shipman}},
  \bibinfo{journal}{Applied Surface Science} \textbf{\bibinfo{volume}{258}},
  \bibinfo{pages}{4161} (\bibinfo{year}{2012}).

\bibitem[{\citenamefont{Bradley}(2012)}]{Bradley2012a}
\bibinfo{author}{\bibfnamefont{R.~M.} \bibnamefont{Bradley}},
  \bibinfo{journal}{Physical Review B} \textbf{\bibinfo{volume}{85}},
  \bibinfo{pages}{115419} (\bibinfo{year}{2012}).

\bibitem[{\citenamefont{Kim et~al.}(2009)\citenamefont{Kim, Joe, Kim, Ha, Lee,
  Kahng, and Kim}}]{Kim2009}
\bibinfo{author}{\bibfnamefont{J.-H.} \bibnamefont{Kim}},
  \bibinfo{author}{\bibfnamefont{M.}~\bibnamefont{Joe}},
  \bibinfo{author}{\bibfnamefont{S.-P.} \bibnamefont{Kim}},
  \bibinfo{author}{\bibfnamefont{N.-B.} \bibnamefont{Ha}},
  \bibinfo{author}{\bibfnamefont{K.-R.} \bibnamefont{Lee}},
  \bibinfo{author}{\bibfnamefont{B.}~\bibnamefont{Kahng}}, \bibnamefont{and}
  \bibinfo{author}{\bibfnamefont{J.-S.} \bibnamefont{Kim}},
  \bibinfo{journal}{Phys. Rev. B} \textbf{\bibinfo{volume}{79}},
  \bibinfo{pages}{205403} (\bibinfo{year}{2009}).

\bibitem[{\citenamefont{Metya and Ghose}(2013)}]{Metya:2013bq}
\bibinfo{author}{\bibfnamefont{A.}~\bibnamefont{Metya}} \bibnamefont{and}
  \bibinfo{author}{\bibfnamefont{D.}~\bibnamefont{Ghose}},
  \bibinfo{journal}{Applied Physics Letters} \textbf{\bibinfo{volume}{103}},
  \bibinfo{pages}{161602} (\bibinfo{year}{2013}).

\bibitem[{\citenamefont{Kim et~al.}(2013)\citenamefont{Kim, Kim,
  Mu{\~n}oz-Garc{\'\i}a, and Cuerno}}]{Kim2013}
\bibinfo{author}{\bibfnamefont{J.-H.} \bibnamefont{Kim}},
  \bibinfo{author}{\bibfnamefont{J.-S.} \bibnamefont{Kim}},
  \bibinfo{author}{\bibfnamefont{J.}~\bibnamefont{Mu{\~n}oz-Garc{\'\i}a}},
  \bibnamefont{and} \bibinfo{author}{\bibfnamefont{R.}~\bibnamefont{Cuerno}},
  \bibinfo{journal}{Physical Review B} \textbf{\bibinfo{volume}{87}},
  \bibinfo{pages}{085438} (\bibinfo{year}{2013}).

\bibitem[{\citenamefont{Redondo-Cubero
  et~al.}(2012)\citenamefont{Redondo-Cubero, Gago, Palomares, M{\"u}cklich,
  Vinnichenko, and V{\'a}zquez}}]{Redondo-cubero2012}
\bibinfo{author}{\bibfnamefont{A.}~\bibnamefont{Redondo-Cubero}},
  \bibinfo{author}{\bibfnamefont{R.}~\bibnamefont{Gago}},
  \bibinfo{author}{\bibfnamefont{F.}~\bibnamefont{Palomares}},
  \bibinfo{author}{\bibfnamefont{A.}~\bibnamefont{M{\"u}cklich}},
  \bibinfo{author}{\bibfnamefont{M.}~\bibnamefont{Vinnichenko}},
  \bibnamefont{and}
  \bibinfo{author}{\bibfnamefont{L.}~\bibnamefont{V{\'a}zquez}},
  \bibinfo{journal}{Physical Review B} \textbf{\bibinfo{volume}{86}},
  \bibinfo{pages}{085436} (\bibinfo{year}{2012}).

\bibitem[{\citenamefont{Hofs\"{a}ss et~al.}(2013)\citenamefont{Hofs\"{a}ss,
  Zhang, Pape, Bobes, and Br\"{o}tzmann}}]{Hofsass2013apa}
\bibinfo{author}{\bibfnamefont{H.}~\bibnamefont{Hofs\"{a}ss}},
  \bibinfo{author}{\bibfnamefont{K.}~\bibnamefont{Zhang}},
  \bibinfo{author}{\bibfnamefont{A.}~\bibnamefont{Pape}},
  \bibinfo{author}{\bibfnamefont{O.}~\bibnamefont{Bobes}}, \bibnamefont{and}
  \bibinfo{author}{\bibfnamefont{M.}~\bibnamefont{Br\"{o}tzmann}},
  \bibinfo{journal}{Applied Physics A} \textbf{\bibinfo{volume}{111}},
  \bibinfo{pages}{653} (\bibinfo{year}{2013}).

\bibitem[{\citenamefont{Valbusa et~al.}(2002)\citenamefont{Valbusa, Boragno,
  and Buatier~de Mongeot}}]{Valbusa2002}
\bibinfo{author}{\bibfnamefont{U.}~\bibnamefont{Valbusa}},
  \bibinfo{author}{\bibfnamefont{C.}~\bibnamefont{Boragno}}, \bibnamefont{and}
  \bibinfo{author}{\bibfnamefont{F.}~\bibnamefont{Buatier~de Mongeot}},
  \bibinfo{journal}{Journal of Physics: Condensed Matter}
  \textbf{\bibinfo{volume}{14}}, \bibinfo{pages}{8153} (\bibinfo{year}{2002}).

\bibitem[{\citenamefont{Bradley and Harper}(1988)}]{Bradley1988}
\bibinfo{author}{\bibfnamefont{R.~M.} \bibnamefont{Bradley}} \bibnamefont{and}
  \bibinfo{author}{\bibfnamefont{J.~M.~E.} \bibnamefont{Harper}},
  \bibinfo{journal}{Journal of Vacuum Science and Technology A}
  \textbf{\bibinfo{volume}{6}}, \bibinfo{pages}{2390} (\bibinfo{year}{1988}).

\bibitem[{\citenamefont{Chan and Chason}(2007)}]{Chan2007}
\bibinfo{author}{\bibfnamefont{W.~L.} \bibnamefont{Chan}} \bibnamefont{and}
  \bibinfo{author}{\bibfnamefont{E.}~\bibnamefont{Chason}},
  \bibinfo{journal}{Journal of Applied Physics} \textbf{\bibinfo{volume}{101}},
  \bibinfo{pages}{121301} (\bibinfo{year}{2007}).

\bibitem[{\citenamefont{Ziegler et~al.}(2010)\citenamefont{Ziegler, Ziegler,
  and Biersack}}]{Ziegler2010}
\bibinfo{author}{\bibfnamefont{J.~F.} \bibnamefont{Ziegler}},
  \bibinfo{author}{\bibfnamefont{M.~D.} \bibnamefont{Ziegler}},
  \bibnamefont{and} \bibinfo{author}{\bibfnamefont{J.~P.}
  \bibnamefont{Biersack}}, \bibinfo{journal}{Nuclear Instruments and Methods in
  Physics Research Section B: Beam Interactions with Materials and Atoms}
  \textbf{\bibinfo{volume}{268}}, \bibinfo{pages}{1818} (\bibinfo{year}{2010}).

\bibitem[{\citenamefont{Carter and Vishnyakov}(1996)}]{Carter1996}
\bibinfo{author}{\bibfnamefont{G.}~\bibnamefont{Carter}} \bibnamefont{and}
  \bibinfo{author}{\bibfnamefont{V.}~\bibnamefont{Vishnyakov}},
  \bibinfo{journal}{Physical Review B} \textbf{\bibinfo{volume}{54}},
  \bibinfo{pages}{17647} (\bibinfo{year}{1996}).

\bibitem[{\citenamefont{Baumann and Schr{\"o}ter}(1991)}]{Baumann1991}
\bibinfo{author}{\bibfnamefont{F.~H.} \bibnamefont{Baumann}} \bibnamefont{and}
  \bibinfo{author}{\bibfnamefont{W.}~\bibnamefont{Schr{\"o}ter}},
  \bibinfo{journal}{Physical Review B} \textbf{\bibinfo{volume}{43}},
  \bibinfo{pages}{6510} (\bibinfo{year}{1991}).

\bibitem[{\citenamefont{Kree et~al.}(2009)\citenamefont{Kree, Yasseri, and
  Hartmann}}]{Kree2009}
\bibinfo{author}{\bibfnamefont{R.}~\bibnamefont{Kree}},
  \bibinfo{author}{\bibfnamefont{T.}~\bibnamefont{Yasseri}}, \bibnamefont{and}
  \bibinfo{author}{\bibfnamefont{A.}~\bibnamefont{Hartmann}},
  \bibinfo{journal}{Nuclear Instruments and Methods in Physics Research Section
  B: Beam Interactions with Materials and Atoms}
  \textbf{\bibinfo{volume}{267}}, \bibinfo{pages}{1403} (\bibinfo{year}{2009}).

\bibitem[{\citenamefont{Shenoy et~al.}(2007)\citenamefont{Shenoy, Chan, and
  Chason}}]{Shenoy2007}
\bibinfo{author}{\bibfnamefont{V.}~\bibnamefont{Shenoy}},
  \bibinfo{author}{\bibfnamefont{W.}~\bibnamefont{Chan}}, \bibnamefont{and}
  \bibinfo{author}{\bibfnamefont{E.}~\bibnamefont{Chason}},
  \bibinfo{journal}{Physical Review Letters} \textbf{\bibinfo{volume}{98}},
  \bibinfo{pages}{256101} (\bibinfo{year}{2007}).

\bibitem[{\citenamefont{Cuerno et~al.}(2011)\citenamefont{Cuerno, Castro,
  Mu{\~n}oz-Garc{\'\i}a, Gago, and V{\'a}zquez}}]{Cuerno2011}
\bibinfo{author}{\bibfnamefont{R.}~\bibnamefont{Cuerno}},
  \bibinfo{author}{\bibfnamefont{M.}~\bibnamefont{Castro}},
  \bibinfo{author}{\bibfnamefont{J.}~\bibnamefont{Mu{\~n}oz-Garc{\'\i}a}},
  \bibinfo{author}{\bibfnamefont{R.}~\bibnamefont{Gago}}, \bibnamefont{and}
  \bibinfo{author}{\bibfnamefont{L.}~\bibnamefont{V{\'a}zquez}},
  \bibinfo{journal}{Nuclear Instruments and Methods in Physics Research Section
  B: Beam Interactions with Materials and Atoms}
  \textbf{\bibinfo{volume}{269}}, \bibinfo{pages}{894} (\bibinfo{year}{2011}).

\bibitem[{\citenamefont{Umbach et~al.}(2001)\citenamefont{Umbach, Headrick, and
  Chan}}]{Umbach2001}
\bibinfo{author}{\bibfnamefont{C.~C.} \bibnamefont{Umbach}},
  \bibinfo{author}{\bibfnamefont{R.~L.} \bibnamefont{Headrick}},
  \bibnamefont{and} \bibinfo{author}{\bibfnamefont{K.-C.} \bibnamefont{Chan}},
  \bibinfo{journal}{Physical Review Letters} \textbf{\bibinfo{volume}{87}},
  \bibinfo{pages}{246104} (\bibinfo{year}{2001}).

\bibitem[{\citenamefont{Castro and Cuerno}(2010)}]{Castro2010}
\bibinfo{author}{\bibfnamefont{M.}~\bibnamefont{Castro}} \bibnamefont{and}
  \bibinfo{author}{\bibfnamefont{R.}~\bibnamefont{Cuerno}},
  \bibinfo{journal}{arXiv preprint arXiv:1007.2144}  (\bibinfo{year}{2010}).

\bibitem[{\citenamefont{Castro and Cuerno}(2012)}]{Castro2012a}
\bibinfo{author}{\bibfnamefont{M.}~\bibnamefont{Castro}} \bibnamefont{and}
  \bibinfo{author}{\bibfnamefont{R.}~\bibnamefont{Cuerno}},
  \bibinfo{journal}{Applied Surface Science} \textbf{\bibinfo{volume}{258}},
  \bibinfo{pages}{4171} (\bibinfo{year}{2012}).

\bibitem[{\citenamefont{Castro et~al.}(2012)\citenamefont{Castro, Gago,
  V{\'a}zquez, Mu{\~n}oz-Garc{\'\i}a, and Cuerno}}]{Castro2012b}
\bibinfo{author}{\bibfnamefont{M.}~\bibnamefont{Castro}},
  \bibinfo{author}{\bibfnamefont{R.}~\bibnamefont{Gago}},
  \bibinfo{author}{\bibfnamefont{L.}~\bibnamefont{V{\'a}zquez}},
  \bibinfo{author}{\bibfnamefont{J.}~\bibnamefont{Mu{\~n}oz-Garc{\'\i}a}},
  \bibnamefont{and} \bibinfo{author}{\bibfnamefont{R.}~\bibnamefont{Cuerno}},
  \bibinfo{journal}{Physical Review B} \textbf{\bibinfo{volume}{86}},
  \bibinfo{pages}{214107} (\bibinfo{year}{2012}).

\bibitem[{\citenamefont{Norris}(2012)}]{Norris2012}
\bibinfo{author}{\bibfnamefont{S.~A.} \bibnamefont{Norris}},
  \bibinfo{journal}{Physical Review B} \textbf{\bibinfo{volume}{86}},
  \bibinfo{pages}{235405} (\bibinfo{year}{2012}).

\bibitem[{\citenamefont{Moreno-Barrado
  et~al.}(2015{\natexlab{a}})\citenamefont{Moreno-Barrado, Castro, Gago,
  V\'azquez, Mu{\~n}oz-Garc{\'\i}a, Redondo-Cubero, Galiana, Ballesteros, and
  Cuerno}}]{Moreno-Barrado2015}
\bibinfo{author}{\bibfnamefont{A.}~\bibnamefont{Moreno-Barrado}},
  \bibinfo{author}{\bibfnamefont{M.}~\bibnamefont{Castro}},
  \bibinfo{author}{\bibfnamefont{R.}~\bibnamefont{Gago}},
  \bibinfo{author}{\bibfnamefont{L.}~\bibnamefont{V\'azquez}},
  \bibinfo{author}{\bibfnamefont{J.}~\bibnamefont{Mu{\~n}oz-Garc{\'\i}a}},
  \bibinfo{author}{\bibfnamefont{A.}~\bibnamefont{Redondo-Cubero}},
  \bibinfo{author}{\bibfnamefont{B.}~\bibnamefont{Galiana}},
  \bibinfo{author}{\bibfnamefont{C.}~\bibnamefont{Ballesteros}},
  \bibnamefont{and} \bibinfo{author}{\bibfnamefont{R.}~\bibnamefont{Cuerno}},
  \bibinfo{journal}{Physical Review B} \textbf{\bibinfo{volume}{91}},
  \bibinfo{pages}{155303} (\bibinfo{year}{2015}{\natexlab{a}}).

\bibitem[{\citenamefont{Moreno-Barrado
  et~al.}(2015{\natexlab{b}})\citenamefont{Moreno-Barrado, Gago,
  Redondo-Cubero, V\'azquez, Mu{\~n}oz-Garc{\'\i}a, Cuerno, Lorenz, and
  Castro}}]{Moreno-Barrado2015b}
\bibinfo{author}{\bibfnamefont{A.}~\bibnamefont{Moreno-Barrado}},
  \bibinfo{author}{\bibfnamefont{R.}~\bibnamefont{Gago}},
  \bibinfo{author}{\bibfnamefont{A.}~\bibnamefont{Redondo-Cubero}},
  \bibinfo{author}{\bibfnamefont{L.}~\bibnamefont{V\'azquez}},
  \bibinfo{author}{\bibfnamefont{J.}~\bibnamefont{Mu{\~n}oz-Garc{\'\i}a}},
  \bibinfo{author}{\bibfnamefont{R.}~\bibnamefont{Cuerno}},
  \bibinfo{author}{\bibfnamefont{K.}~\bibnamefont{Lorenz}}, \bibnamefont{and}
  \bibinfo{author}{\bibfnamefont{M.}~\bibnamefont{Castro}},
  \bibinfo{journal}{EPL} \textbf{\bibinfo{volume}{109}}, \bibinfo{pages}{48003}
  (\bibinfo{year}{2015}{\natexlab{b}}).

\bibitem[{\citenamefont{Moreno-Barrado
  et~al.}(2015{\natexlab{c}})\citenamefont{Moreno-Barrado, Castro,
  Mu{\~n}oz-Garc{\'\i}a, and Cuerno}}]{Moreno-Barrado2015c}
\bibinfo{author}{\bibfnamefont{A.}~\bibnamefont{Moreno-Barrado}},
  \bibinfo{author}{\bibfnamefont{M.}~\bibnamefont{Castro}},
  \bibinfo{author}{\bibfnamefont{J.}~\bibnamefont{Mu{\~n}oz-Garc{\'\i}a}},
  \bibnamefont{and} \bibinfo{author}{\bibfnamefont{R.}~\bibnamefont{Cuerno}},
  \bibinfo{journal}{Nuclear Instruments and Methods in Physics Research Section
  B}  (\bibinfo{year}{2015}{\natexlab{c}}).

\bibitem[{\citenamefont{Cross and Greenside}(2009)}]{Cross2009}
\bibinfo{author}{\bibfnamefont{M.}~\bibnamefont{Cross}} \bibnamefont{and}
  \bibinfo{author}{\bibfnamefont{H.}~\bibnamefont{Greenside}},
  \emph{\bibinfo{title}{Pattern Formation and Dynamics in Nonequilibrium
  Systems}} (\bibinfo{publisher}{Cambridge University Press},
  \bibinfo{address}{Cambridge, England}, \bibinfo{year}{2009}).

\bibitem[{\citenamefont{Norris}(2013)}]{Norris2013b}
\bibinfo{author}{\bibfnamefont{S.~A.} \bibnamefont{Norris}},
  \bibinfo{journal}{Journal of Applied Physics} \textbf{\bibinfo{volume}{114}},
  \bibinfo{pages}{204303} (\bibinfo{year}{2013}).

\bibitem[{\citenamefont{Hofs{\"a}ss}(2015)}]{Hofsass2015}
\bibinfo{author}{\bibfnamefont{H.}~\bibnamefont{Hofs{\"a}ss}},
  \bibinfo{journal}{unpublished}  (\bibinfo{year}{2015}).

\bibitem[{\citenamefont{Zhou et~al.}(2007)\citenamefont{Zhou, Wang, Zhou,
  Headrick, {\"O}zcan, Wang, {\"O}zaydin, Jr., and Siddons}}]{Zhou2007}
\bibinfo{author}{\bibfnamefont{H.}~\bibnamefont{Zhou}},
  \bibinfo{author}{\bibfnamefont{Y.}~\bibnamefont{Wang}},
  \bibinfo{author}{\bibfnamefont{L.}~\bibnamefont{Zhou}},
  \bibinfo{author}{\bibfnamefont{R.~L.} \bibnamefont{Headrick}},
  \bibinfo{author}{\bibfnamefont{A.~S.} \bibnamefont{{\"O}zcan}},
  \bibinfo{author}{\bibfnamefont{Y.}~\bibnamefont{Wang}},
  \bibinfo{author}{\bibfnamefont{G.}~\bibnamefont{{\"O}zaydin}},
  \bibinfo{author}{\bibfnamefont{K.~F.~L.} \bibnamefont{Jr.}},
  \bibnamefont{and} \bibinfo{author}{\bibfnamefont{D.~P.}
  \bibnamefont{Siddons}}, \bibinfo{journal}{Physical Review B}
  \textbf{\bibinfo{volume}{75}}, \bibinfo{pages}{155416}
  (\bibinfo{year}{2007}).

\bibitem[{\citenamefont{Madi et~al.}(2008)\citenamefont{Madi, Davidovitch,
  George, Norris, Brenner, and Aziz}}]{Madi2008}
\bibinfo{author}{\bibfnamefont{C.}~\bibnamefont{Madi}},
  \bibinfo{author}{\bibfnamefont{B.}~\bibnamefont{Davidovitch}},
  \bibinfo{author}{\bibfnamefont{H.}~\bibnamefont{George}},
  \bibinfo{author}{\bibfnamefont{S.}~\bibnamefont{Norris}},
  \bibinfo{author}{\bibfnamefont{M.}~\bibnamefont{Brenner}}, \bibnamefont{and}
  \bibinfo{author}{\bibfnamefont{M.}~\bibnamefont{Aziz}},
  \bibinfo{journal}{Physical Review Letters} \textbf{\bibinfo{volume}{101}},
  \bibinfo{pages}{246102} (\bibinfo{year}{2008}).

\bibitem[{\citenamefont{Madi et~al.}(2009)\citenamefont{Madi, George, and
  Aziz}}]{Madi2009}
\bibinfo{author}{\bibfnamefont{C.~S.} \bibnamefont{Madi}},
  \bibinfo{author}{\bibfnamefont{H.~B.} \bibnamefont{George}},
  \bibnamefont{and} \bibinfo{author}{\bibfnamefont{M.}~\bibnamefont{Aziz}},
  \bibinfo{journal}{Journal of Physics: Condensed Matter}
  \textbf{\bibinfo{volume}{21}}, \bibinfo{pages}{224010}
  (\bibinfo{year}{2009}).

\bibitem[{\citenamefont{Madi and Aziz}(2012)}]{Madi2012}
\bibinfo{author}{\bibfnamefont{C.~S.} \bibnamefont{Madi}} \bibnamefont{and}
  \bibinfo{author}{\bibfnamefont{M.~J.} \bibnamefont{Aziz}},
  \bibinfo{journal}{Applied Surface Science} \textbf{\bibinfo{volume}{258}},
  \bibinfo{pages}{4112} (\bibinfo{year}{2012}).

\end{thebibliography}

\end{document}